\documentclass[prd,nofootinbib,preprint,superscriptaddress,floatfix]{revtex4}

\usepackage{amsmath}
\usepackage[dvips]{graphicx}

\newcommand{\capdef}{}
\newcommand{\mycaption}[2][\capdef]{\renewcommand{\capdef}{#2}%
       \caption[#1]{{\footnotesize #2}}}
\makeatletter
\renewcommand{\fnum@table}{\textbf{\tablename~\thetable}}
\renewcommand{\fnum@figure}{\textbf{\figurename~\thefigure}}
\makeatother

\newcommand{\Dmq}{\Delta m^2}

\newcommand{\anu}{\overline{\nu}}
\newcommand{\delCP}{\delta_\mathrm{CP}}
\newcommand{\stheta}{\sin^22\theta_{13}}
\newcommand{\ND}{\mathrm{ND}}
\newcommand{\FD}{\mathrm{FD}}

\begin{document}


\vspace*{10mm}

\title{On the impact of systematical uncertainties for the CP violation
measurement in superbeam experiments}

\author{Patrick Huber}
\email{pahuber_at_vt.edu}
\affiliation{Physics Department, Theory Division, CERN, CH--1211
  Geneva 23, Switzerland}
\affiliation{Institute for Particle, Nuclear and Astronomical
  Sciences,\\
Physics Department, Virgina Tech, Blacksburg, VA 24062, USA}

\author{Mauro Mezzetto} 
\email{mauro.mezzetto_at_pd.infn.it}
\affiliation{Instituto Nazionale Fisica Nucleare, Sezione di Padova,
  Via Marzolo 8, 35100 Padova, Italy}

\author{Thomas Schwetz} 
\email{schwetz_at_cern.ch}
\affiliation{Physics Department, Theory Division, CERN, CH--1211
  Geneva 23, Switzerland}

\vspace*{.5cm}

\begin{abstract}
  \vspace*{.5cm} Superbeam experiments can, in principle, achieve
  impressive sensitivities for CP violation in neutrino oscillations
  for large $\theta_{13}$. We study how those sensitivities depend on
  assumptions about systematical uncertainties. We focus on the second
  phase of T2K, the so-called T2HK experiment, and we explicitly
  include a near detector in the analysis. Our main result is that
  even an idealised near detector cannot remove the dependence on
  systematical uncertainties completely.  Thus additional information
  is required.  We identify certain combinations of uncertainties,
  which are the key to improve the sensitivity to CP violation, for
  example the ratio of electron to muon neutrino cross sections and
  efficiencies. For uncertainties on this ratio larger than 2\%, T2HK
  is systematics dominated. We briefly discuss how our results apply
  to a possible two far detector configuration, called T2KK. We do not
  find a significant advantage with respect to the reduction of
  systematical errors for the measurement of CP violation for this
  setup.
\end{abstract}

\preprint{CERN-PH-TH/2007-227}
\preprint{VPI-IPNAS-07-09}

\maketitle


\section{Introduction}

Neutrino oscillation offers a natural, minimalistic framework to
account for the observed data in a wide variety of experiments. It has
been established as the leading mechanism for flavour transitions in
solar~\cite{Hosaka:2005um, Aharmim:2005gt} and
atmospheric~\cite{Ashie:2005ik} neutrinos. The typical $L/E$ pattern
expected for oscillations begins to emerge from various data sets like
the most recent KamLAND data~\cite{nnn07, Araki:2004mb} or the first
generation of long-baseline $\nu_\mu$ disappearance experiments
K2K~\cite{Ahn:2006zz, Aliu:2004sq} and MINOS~\cite{Michael:2006rx}.
The only experiment which cannot be accounted for
in a three flavour oscillation framework is the LSND $\anu_e$
appearance signal~\cite{Aguilar:2001ty}. The LSND result is quite
difficult to reconcile with the null results of
CDHS~\cite{Dydak:1983zq} and Bugey~\cite{Declais:1994su}, even when
one allows for one or more sterile neutrinos. Recently, MiniBooNE
failed to confirm the LSND signal~\cite{AguilarArevalo:2007it}. For 
an analysis of sterile neutrino solutions to the LSND result in view
of the MiniBooNE data see~\cite{Maltoni:2007zf}. The status of LSND
thus remains unclear and therefore we will assume that LSND has a
non-oscillation explanation. We will consider only oscillations
between three active flavours.

Neutrino oscillations require massive neutrinos. This on its own is
one of the strongest and first indications for physics beyond the
Standard Model. Most models for neutrino mass generation point to a
very high energy scale far beyond the reach of current and future
accelerator experiments. The neutrino is thus a unique messenger for
otherwise in-accessible physics. To fully exploit this potential, new,
high precision oscillation experiments are necessary. These
experiments will address the size of $\theta_{13}$, the neutrino mass
hierarchy, leptonic CP violation and whether $\theta_{23}=\pi/4$. None
of the currently running or approved experiments has sufficient
sensitivity to achieve an accurate measurement of the neutrino mass
hierarchy or CP violation~\cite{Huber:2004ug}. The reason is, that in
both cases the effects are quite small and subtle. There is a plethora
of possible technologies to address these questions. They range from
third generation superbeam experiments to  beta beams and neutrino
factories. For a recent review see~\cite{Group:2007kx}.

In this paper we focus on third generation superbeam experiments.
These experiments are based on conventional neutrino beams from
$\pi$-decay. The experiments are `super' in the sense that they will
use proton beams of unprecedented strength around $1-4\,\mathrm{MW}$
and detectors with fiducial masses of several $100\,\mathrm{kt}$. This
will allow to collect many thousands of $\nu_e$ and $\anu_e$
appearance events (assuming $\sin^22\theta_{13}=0.1$). Thus
statistical errors will be at most of percent size and systematical
errors will be no longer be negligible. Nearly all previously
published sensitivity studies use some {\it ad-hoc}\footnote{{\it
    ad-hoc} refers to the fact that this value is usually smaller than
  any value achieved by any previous experiment.}  value of
systematics which is assumed to be achieved by means of a near
detector.  This near detector is not specified nor included in the
calculation.

In this paper we carefully investigate a large number of possible
contributions to the systematical error budget in superbeam
experiments. To be specific we use the T2K~\cite{Itow:2001ee}
experiment as example, focusing mainly on the discovery of CP
violation in T2K phase~II (T2HK). We briefly comment also on T2K
phase~I, as well as the T2KK
proposal~\cite{Ishitsuka:2005qi,Hagiwara:2005pe}, where half of the
T2HK detector is moved to Korea.  Our analysis is based on a realistic
Monte Carlo study of the detector response and we explicitly include a
(though somewhat idealised) near detector in the simulation. The goal
of this paper is to investigate to which extent a near detector can
contribute to reduce the impact of various systematical error sources.
We will show that even an idealised near detector on its own cannot
reduce the impact of all error terms. Thus additional information on
fluxes and/or cross sections will be required.

The outline of the paper is as follows. In Sec.~\ref{sec:qualitative}
we discuss qualitatively the impact of systematics on an appearance
experiment. Though highly simplified this discussion will allow us to
understand many features of the numerical calculations. In
Sec.~\ref{sec:xsec} we review in some detail uncertainties on neutrino
cross sections, since they will be crucial in the subsequent
analysis. In Sec.~\ref{sec:simulation} we give a brief description of
our experiment simulation and the various types of systematics we
include in our analysis. More technical details can be found in
appendix~\ref{app:simulation}. Sec.~\ref{sec:results} contains the
results of our numerical calculations: in Sec.~\ref{sec:CPV} we
consider the CP violation sensitivity of T2HK discussing mainly the
impact of cross section uncertainties, whereas in
Sec.~\ref{sec:fluxes-ND} we point out under which circumstances
information on fluxes and an improved near detector are useful. In
Sec.~\ref{sec:th13} we consider the determination of $\theta_{13}$ in
the context of T2K (phase~I) and T2HK (phase~II), whereas in
Sec.~\ref{sec:t2kk} we discuss the systematics impact for T2KK in
comparison with T2HK focusing again on CP violation. A summary and
some speculative remarks on other high precision oscillation
facilities follow in Sec.~\ref{sec:conclusions}. In
appendix~\ref{app:detector} we give technical details on the
experiment simulation, whereas in appendix~\ref{app:syst} we describe
the statistical analysis including the implementation of systematics.
In appendix~\ref{app:effectiveFD} we show how the full ND/FD
simulation can be approximated by a much simpler FD-only analysis
adopting only very few ``effective'' systematics parameters.

\section{Qualitative discussion}
\label{sec:qualitative}

Before we plunge into a detailed numerical study, we will present here
a simple argument, why even an idealised near detector is not
sufficient for an appearance experiment. For the sake of discussion we
introduce a couple of simplifications, for example we consider only
total rates and neglect some background sources, while we do include
them in the subsequent numerical calculations, see
Sec.~\ref{sec:simulation} and App.~\ref{app:simulation} for details.

The total number of $\nu_e$ and $\nu_\mu$ events in near detector (ND)
and far detector (FD) can be written as
\begin{eqnarray}
n_{\nu_\mu}^\ND &=& \frac{N_\ND}{L_\ND^2} \, 
                     \Phi_{\nu_\mu} \, \sigma_{\nu_\mu}\epsilon_{\nu_\mu} 
\label{eq:NDm}\\
n_{\nu_e}^\ND   &=& \frac{N_\ND}{L_\ND^2} \left[  
                     \Phi_{\nu_e} \, \sigma_{\nu_e}\epsilon_{\nu_e} +
                     n_\mathrm{NC}^\ND \right]
\label{eq:NDe}\\
n_{\nu_\mu}^\FD &=& \frac{N_\FD}{L_\FD^2} \,
                     \Phi_{\nu_\mu} \, P(\nu_\mu\to\nu_\mu) \, 
                     \sigma_{\nu_\mu}\epsilon_{\nu_\mu} 
\label{eq:FDm}\\
n_{\nu_e}^\FD   &=& n_{\nu_e}^\mathrm{FD,sig} + n_{\nu_e}^\mathrm{FD,bg} 
\label{eq:FDe}
\end{eqnarray}
with 
\begin{eqnarray}
  n_{\nu_e}^\mathrm{FD,sig} &=& \frac{N_\FD}{L_\FD^2} 
   \Phi_{\nu_\mu} \, P(\nu_\mu\to\nu_e) \, 
   \sigma_{\nu_e}\epsilon_{\nu_e} \,,
\label{eq:FDsig}\\
  n_{\nu_e}^\mathrm{FD,bg} &=& \frac{N_\FD}{L_\FD^2} \left[ 
                     \Phi_{\nu_e} \, P(\nu_e\to\nu_e) \, 
                     \sigma_{\nu_e}\epsilon_{\nu_e}  +
                     n_\mathrm{NC}^\FD \right] \,.
\label{eq:FDbg}
\end{eqnarray}
Here $N$ is the total normalisation (number of target nuclei),
$\sigma_{\nu_\alpha}$ is the charged current cross section for
$\nu_\alpha$, $\epsilon_{\nu_\alpha}$ is the detection efficiency for
$\nu_\alpha$ (assumed to be identical for ND and FD),
$P(\nu_\beta\rightarrow\nu_\alpha)$ is the probability for a neutrino
of flavour $\beta$ to oscillate into flavour $\alpha$,
$\Phi_{\nu_\beta}$ is the initial neutrino flux, and $L$ is the
distance from the detector to the source. For the $\nu_e$ signal we
include here the intrinsic $\nu_e$ beam contamination and the
background from neutral current (NC) interactions $n_\mathrm{NC}
\times N/L^2$, whereas for the disappearance channel we neglect
backgrounds. Note, that the efficiency $\epsilon$ and the cross
section $\sigma$ appear as product and hence we define an effective
cross section
\begin{equation}
\tilde\sigma_{\nu_\alpha}:=\sigma_{\nu_\alpha}\epsilon_{\nu_\alpha}\,.
\end{equation}

Many of the quantities---most importantly cross sections and
fluxes---appearing in Eqs.~(\ref{eq:NDm}) to (\ref{eq:FDbg}) are
subject to (sometimes large) uncertainties. Therefore, the aim is to
use data from the ND in order to predict the signal in the FD,
reducing the dependence on external information as much as possible.
This can be done efficiently for a disappearance measurement. Using
Eqs.~(\ref{eq:NDm}) and (\ref{eq:FDm}) one finds
\begin{equation}\label{eq:dis}
n_{\nu_\mu}^\FD = n_{\nu_\mu}^\ND \,
                   \frac{N_\FD}{N_\ND} \frac{L_\ND^2}{L_\FD^2} \, 
                   \, P(\nu_\mu\to\nu_\mu) \,.
\end{equation}
Hence, assuming that the uncertainty on $N_\FD / N_\ND \times L_\ND^2
/ L_\FD^2$ is negligible, a complete cancellation of all systematical
errors happens (in this idealised discussion), since the same
combination of $\tilde\sigma$ and $\Phi$ appears in ND and FD. This
result is well known and has been exploited with success in the
K2K~\cite{Ahn:2006zz} and MINOS~\cite{Michael:2006rx}
experiments.\footnote{Certainly, in real life the situation is much
more complicated than suggested by Eq.~(\ref{eq:dis}) which should
illustrate the principle. In the K2K and MINOS analyses many
additional complications have been taken into account. For example, in
both cases the fluxes are rather different in ND and FD, and effects of 
the energy spectrum are included via a near-to-far extrapolation matrix.}
Also, it is the basis for all of the latest generation of reactor
neutrino experiments like
DoubleChooz~\cite{Huber:2003pm,Ardellier:2006mn}. For reactor
experiments, the idea is that by careful design and construction, the
near to far comparison can be made so precise that Eq.~(\ref{eq:dis})
holds to an accuracy of better than 1\%.\footnote{For reactor
experiments, this very high accuracy is possible because the fiducial
mass will be determined within less than 0.5\%. Moreover, the
efficiencies are close to 100\%, hence there will be only a very small
error due to event reconstruction.}

However, the situation is very different for an appearance experiment.
Depending on the relative importance of the two terms in
Eq.~(\ref{eq:FDe}) we can identify two qualitatively different regimes
for the appearance measurement: first, the regime close to the
sensitivity limit of the experiment ({\it i.e.}, small $\stheta$),
where the $\nu_e$ events are dominated by background, and second, the
regime of large $\stheta$, where the actual appearance signal
dominates over the background in Eq.~(\ref{eq:FDe}). Hence, depending
on the regime, one expects that either the error on the background or
on the signal is most relevant for the sensitivity. As the numerical
calculations will show, for T2HK the transition between the two
regimes occurs roughly at $\stheta \simeq 0.01$.

In the background dominated case of small $\stheta$ the ND plays a
crucial role in measuring the background. Under the assumption that
the NC background is the same in ND and FD, and neglecting the small
effect of oscillations on the beam background, $P(\nu_e\to\nu_e)
\approx 1$, the background in the FD can be predicted by the $\nu_e$
events in the ND from Eq.~(\ref{eq:NDe}):
\begin{equation}\label{eq:app-bg}
  n_{\nu_e}^\mathrm{FD,bg} = n_{\nu_e}^\ND \,
  \frac{N_\FD}{N_\ND} \frac{L_\ND^2}{L_\FD^2}  \,.
\end{equation}
It becomes important how well the above assumptions are fulfilled, for
example, how well one can extrapolate the NC background from ND to
FD. Moreover, the statistical precision for $n_{\nu_e}^\ND$ is an
issue, {\it i.e.}, the size of the ND, since the beam contains only a
small component of $\nu_e$, at the level of 1\% of $\nu_\mu$. 
Note that in the discussion leading to Eq.~(\ref{eq:app-bg}) we have
neglected a background coming from $\nu_\mu$ charged-current
interactions. This background is very different in ND and FD because
of oscillations of $\nu_\mu$ with $\sin^22\theta_{23} \simeq
1$. Therefore, this additional background component will further
complicate the extrapolation of the background measurement from the ND
to the FD. While we neglect such a background in the current section
for simplicity, we do include a background from $\mu/e$
miss-identification in the numerical calculations, see
Sec.~\ref{sec:simulation}.

The signal dominated regime of large $\stheta$ is probably the more
interesting case, since it will allow for high precision measurements
such as CP violation (CPV), and therefore the main focus of our work
is on this case. From Eq.~(\ref{eq:FDsig}) one can see that for the
signal the combination $\Phi_{\nu_\mu} \times \tilde\sigma_{\nu_e}$ is
relevant, which cannot be determined by the ND, and
Eqs.~(\ref{eq:NDm}) and (\ref{eq:FDsig}) combine to
\begin{equation}\label{eq:app-sig}
n_{\nu_e}^\mathrm{FD,sig} = n_{\nu_\mu}^\ND \,
     \frac{N_\FD}{N_\ND} \frac{L_\ND^2}{L_\FD^2} \, 
     \frac{\tilde\sigma_{\nu_e}}{\tilde\sigma_{\nu_\mu}} \,
     P(\nu_\mu\to\nu_e) \,.
\end{equation}
Obviously, there remains some dependence on the effective cross
sections, namely the ratio
$\tilde\sigma_{\nu_e}/\tilde\sigma_{\nu_\mu}$ survives. The ability
to discover CPV largely depends on the ability to compare the neutrino
and anti-neutrino appearance signals, thus it is useful to look at the
ratio of the corresponding event rates:
\begin{equation}\label{eq:CP-ratio}
  \frac{n_{\nu_e}^\mathrm{FD,sig}}{n_{\anu_e}^\mathrm{FD,sig}} = 
  \frac{n_{\nu_\mu}^\ND}{n_{\anu_\mu}^\ND} \, 
  \frac{\tilde\sigma_{\nu_e}}{\tilde\sigma_{\nu_\mu}} \,
  \frac{\tilde\sigma_{\anu_\mu}}{\tilde\sigma_{\anu_e}} \,
  \frac{P(\nu_\mu\rightarrow\nu_e)}{P(\anu_\mu\rightarrow\anu_e)}\,.
\end{equation}
From this discussion we learn that one of the following combinations
of quantities has to be known in order to predict the signal for
the CPV measurement:
\begin{equation}\label{eq:combinations}
  \left(\frac{\tilde\sigma_{\nu_e}}{\tilde\sigma_{\nu_\mu}} \,,
  \frac{\tilde\sigma_{\anu_e}}{\tilde\sigma_{\anu_\mu}} \right)
  \quad\text{or}\quad
  \left(\frac{\tilde\sigma_{\nu_e}}{\tilde\sigma_{\anu_e}} \,,
  \frac{\tilde\sigma_{\nu_\mu}}{\tilde\sigma_{\anu_\mu}} \right)
  \quad\text{or}\quad
  \left(\frac{\Phi_{\nu_\mu}}{\Phi_{\anu_\mu}} \,,
  \frac{\tilde\sigma_{\nu_e}}{\tilde\sigma_{\anu_e}} \right) \,.
\end{equation}
The first two combinations follow from Eq.~(\ref{eq:CP-ratio}): if
either the flavour ratio of effective cross sections for neutrino and
anti-neutrinos separately or the neutrino/anti-neutrino ratio for
$\nu_e$ and $\nu_\mu$ separately are known with good precision then
the high statistics $\nu_\mu$ and $\anu_\mu$ samples from the ND allow
to predict the CPV signal in the FD. Note that this does not require
knowledge on the double ratio
$(\tilde\sigma_{\nu_e}/\tilde\sigma_{\nu_\mu})/
(\tilde\sigma_{\anu_e}/\tilde\sigma_{\anu_\mu})$.  The last
combination in Eq.~(\ref{eq:combinations}) follows directly from
Eq.~(\ref{eq:FDsig}): If $\nu_\mu$ and $\anu_\mu$ fluxes, as well as
$\nu_e$ and $\anu_e$ effective cross sections are known the signal can
directly be predicted without the need of the ND.
Let us mention again that in Eq.~(\ref{eq:combinations}) always the
product of cross sections times efficiencies appears. Uncertainties on
both of them contribute to the error on $\tilde\sigma$.
Although the preceding discussion is highly simplified it captures
quite well the behaviour of the near/far detector system for an
appearance experiment. Many results of our numerical simulation can be
understood qualitatively with this kind of reasoning, and in the
course of our discussion we will refer frequently to the arguments
presented in this section. In appendix~\ref{app:effectiveFD} we
demonstrate that these arguments can be used to approximate the full
ND/FD simulation by a rather simple (in what concerns the systematics
treatment) FD-only setup.

\section{Neutrino cross sections}
\label{sec:xsec}

Since cross section ratios play such an important role in the
discussion of systematics, we would like to make a few remarks. Based
purely on existing experimental data~\cite{PDG}, without the use of a
specific model, the errors are in the range $20-50\%$. Especially
anti-neutrino cross sections are not well measured or for some energy
ranges not measured at all. T2HK operates in the energy range from
$400-1200\,\mathrm{MeV}$, therefore most events (at least after the
single ring cut) will be due to quasi-elastic (QE) reactions and we
will focus on these. The theory for neutrino scattering off a free
nucleon is well understood~\cite{LlewellynSmith:1971zm}. Here the
cross section is given as a function of various form
factors\footnote{There are some deviations from pure dipole form,
  which, however, can be accounted for, see {\it
    e.g.}~\cite{Bodek:2007vi}.}, which are all but one well measured.
The one `free' parameter is the axial mass $m_A$.  Based on that
formulation, one would expect that the ratio of $\sigma_{\nu_e}$ to
$\sigma_{\nu_\mu}$ can be accurately computed.  However, in most
detectors the bulk of the fiducial mass stems from heavier nuclei,
therefore nuclear effects are non-negligible.  Many experiments use
the Smith-Moniz formalism~\cite{Smith:1972xh} to account for nuclear
effects. Here, the nucleus is described as a Fermi gas of nucleons and
two new parameters enter: the Fermi momentum $k_F$ and the binding
energy $E_B$. This description thus introduces a smearing of the
energy of the outgoing neutrino due to Fermi motion and a reduction of
the cross section due to Pauli blocking. There are basically two
differences between electron and muon neutrino scattering: one is
the pure kinematic effect of the difference in $m_e$ and $m_\mu$.
This effect is trivial to account for. The other one is, the fact the momentum
transfer to the nucleus is going to be different. In order to compute
the resulting effect on the ratio it is necessary to know the momentum
distribution of the bound nucleon.

Moreover, it is not clear how well the Fermi gas model actually captures the
physics of the interaction of a neutrino with bound nucleons. Until
recently the consensus value for the axial mass was
$m_A=1.025\pm0.021\,\mathrm{GeV}$~\cite{Bernard:2001rs}. This is
somewhat in contrast with values reported by K2K~\cite{Gran:2006jn} of
$m_A=1.20\pm0.12\,\mathrm{GeV}$ and MiniBooNE~\cite{MB:2007ru} of
$m_A=1.23\pm0.20\,\mathrm{GeV}$. One possible explanation, put forth
in~\cite{MB:2007ru}, is that the old data was mainly obtained on
Deuterium, where nuclear effects are very small, whereas the new data
used Oxygen (K2K) or Carbon (MiniBooNE). This in turn would indicate
that there are nuclear effects which are not properly included in the
Smith-Moniz formalism and thus the value of $m_A$ determined from
nuclei is in reality an effective parameter. However, it was
pointed out in~\cite{Bodek:2007vi} that $m_A$ should be the same or
decrease in a nuclear target compared to deuterium, see
also~\cite{Singh:1992dc}. Thus the experimental situation on the
quasi-elastic cross section itself is somewhat unclear. It may be
that the data on which the value for $m_A$ in~\cite{Bernard:2001rs} is
based are not pure QE events, which of course would introduce a bias
into the determination of of $m_A$. The value of $m_A$ itself is not
expected to have a large impact on the ratio of cross sections. This
example is intended to show that there are still open issues in
seemingly well understood neutrino interaction processes. Also a
comparison of several state of art event generator in the range
$0<E_\nu<2\,\mathrm{GeV}$ yields errors in the range
5\%--15\%~\cite{Monroe:2004xe}.

Clearly, from a purely theoretical point of view any correction due to
finite lepton masses should be small especially for energies $E\gg
m_\mu$, which is the case for T2HK. Surprisingly we found only very
little literature discussing the cross section ratio.  There are many
papers computing either the $\nu_\mu$ or $\nu_e$ QE cross section on
various nuclei, but only in Refs.~\cite{Kolbe:2003ys, Valverde:2006zn,
Benhar:2006nr} we could find a result for both, $\nu_\mu$ and $\nu_e$ on
Oxygen\footnote{We do not claim that our survey is complete, but it
certainly is representative of the small number of pertinent results
compared to the total amount of literature about neutrino cross
sections.}. In Ref.~\cite{Valverde:2006zn} the error on the ratio is
quantified explicitly, however only in the energy range below
$500\,\mathrm{MeV}$. Their result is, that within the model given
in~\cite{Nieves:2004wx} the errors are about 1\% coming from the
uncertainties of the input parameters. They also estimate that
physical effects not accounted for introduce no more than 5\% error on
the ratio. Ref.~\cite{Benhar:2006nr} gives both cross sections in the
energy range relevant for T2HK. The authors of~\cite{Benhar:2006nr}
kindly provided their results and we could compute the ratio and
compare it to the results in~\cite{Nieves:2004wx}. At
$450\,\mathrm{MeV}$ we find a difference in the ratios of about 3\%,
which falls within the error estimate given in~\cite{Nieves:2004wx}.

We furthermore compared the results for the ratio obtained with the
event generators NUANCE~\cite{Nuance}, GENIE~\cite{genie} and
NEUGEN~\cite{neugen} and we find between $400-1200\,\mathrm{MeV}$ a
spread of about $1\%$\footnote{NEUGEN and GENIE give nearly identical
results.}. With respect to the theoretical calculation
in~\cite{Benhar:2006nr} we find a difference of $3\%$ at
$400\,\mathrm{MeV}$ which decreases down to $0.5\%$ at
$1200\,\mathrm{MeV}$. We extracted the data
in~\cite{Kolbe:2003ys,Valverde:2006zn} from the published plots and,
within the errors this inevitably introduces, they show a similar
spread. Summarising, all theoretical sources for the ratio we could
find, showed a spread of $3\%$ or less throughout the energy range
relevant for T2HK. Let us mention that in the energy region of a few
$100\,\mathrm{MeV}$, the error on the ratio from theory reaches more
than 10\%. This energy range is relevant for a beta beam with a
Lorentz-$\gamma$ of around 100 or the SPL
superbeam~\cite{Campagne:2006yx}. Therefore, in these cases present
theory calculations do not provide a relevant constraint on the ratio.
Also note, that the spread for the ratio of neutrino to anti-neutrino
cross sections is found to be larger than 10\% throughout the whole
energy range for T2HK. 

MiniBooNE offers an excellent case study of how a recent experiment
deals with these issues. MiniBooNE is a $\nu_e$ appearance experiment
in the same energy region than T2HK, which uses, instead of a near
detector, the unoscillated $\nu_\mu$ sample in the far detector to
predict their $\nu_e$ backgrounds and also the $\nu_e$ signal. A
dedicated investigation of the effect of cross section uncertainties
has been performed, and the results of Ref.~\cite{Monroe:2006pi}
indicate that the $\nu_\mu$ to $\nu_e$ cross section ratio has an
error of about $8-9\%$.

The question is, whether one trusts theory calculations, which state
that the ratio is known to better than 3\%. If this is the case, the
cross section ratio would have only a small impact on the overall
error budget and the near-far comparison would effectively control the
systematics also in an appearance experiment. On the other hand, one
should acknowledge the fact that neutrino scattering data is sparse
and no theory of quasi-elastic scattering has been experimentally
tested with a large degree of accuracy. Thus there is, at least in
principle, considerable room for surprises and consensus is needed on
whether this risk is tolerable in view of a large scale project such
as considered here. For instance, there is a long standing excess of
$\nu_e$ events in sub-GeV atmospheric neutrino
data~\cite{Ashie:2005ik}, whose origin so far is not understood and
might reflect a problem in the $\nu_e/\nu_\mu$ cross section
ratio.\footnote{We thank E.~Lisi for pointing this out.} Also, the
error estimate obtained by MiniBooNE clearly points towards much
larger errors of the order $10\%$, than our survey of theory results
has found.

Summarising, it seems that the 10\% default errors used here for the
individual cross sections are somewhat optimistic, especially if one
keeps in mind that basically all existing data is nearly exclusively
for $\nu_\mu$. On the other hand, with our defaults and assumption of
uncorrelated errors for all cross sections, the ratio
$\sigma_{\nu_e}/\sigma_{\nu_\mu}$ would have an effective error of
$\sim10\%$ (see appendix~\ref{app:effectiveFD}), which may be larger
than expected from theory but is in agreement with the MiniBooNE
numbers. In the following we will take the conservative view point,
that we would like to be independent of theoretical arguments about
the cross section ratio and use (by then) available experimental data
to control systematical uncertainties. Nonetheless, we show results
for various constraints on this ratio. Note, that the error relevant
for the oscillation analysis is the error on the ratio of the product
of cross section and efficiency. Thus, even if there is a tight
constraint on the cross section ratios, the efficiencies still will
need to be determined accurately by other means.

\section{Description of the simulation}
\label{sec:simulation}

Let us now give some key features of our numerical simulation. More
technical details are deferred to appendix~\ref{app:simulation}.  We
consider the following standard setup for the phase~I (phase~II) T2K
(T2HK) configuration. The fiducial far detector mass is 22.5~kt
(500~kt) and the beam power is 0.77~MW (4~MW). The running time is for
T2K and T2HK 2~yr for the neutrino and 6~yr for the anti-neutrino
beams. The baseline is $295\,\mathrm{km}$ and we use an average matter
density of $2.8\,\mathrm{g}\,\mathrm{cm}^{-3}$. This setup is based
on~\cite{Itow:2001ee}. Details of the T2KK setup are given in
Sec.~\ref{sec:t2kk}.
We consider the signals from $\nu_\mu$ and $\nu_e$ single ring events,
for both the neutrino and anti-neutrino run, {\it i.e.}, disappearance
and appearance channels, and we take into account also the effect of
oscillations on various background components. Energy resolution is
treated with migration matrices including nuclear effects as well as
the contamination of the single ring sample with non-quasi-elastic
events.  We include the intrinsic $\nu_e$, $\anu_e$ and $\anu_\mu$
(and the CP conjugate ones for the anti-neutrino run) as well as
neutral current backgrounds. We restrict the analysis to reconstructed
neutrino energies from $0.4-1.2\,\mathrm{GeV}$. This range is divided
into 8 equidistant bins.  For the near detector we assume a water
\v{C}erenkov detector with fiducial mass of $0.1\,\mathrm{kt}$ and
otherwise identical properties to the far detector. Specifically, we
also assume the same acceptance, {\it i.e.}\ a flat near-far
ratio. This assumption translates into the requirement that the near
detector distance must be large enough in order to see the decay pipe
as point source like the far detector.  To be specific, we follow the
choice of the T2K collaboration and use a baseline of $2\,\mathrm{km}$
for the near detector~\cite{2kmLOI}.  Further details on the detector
simulation are to be found in appendix~\ref{app:simulation}.

The $\chi^2$ computation is based on a standard Poissonian form and we
use the so-called pull approach~\cite{Huber:2002mx, Fogli:2002pt} to
include the various sources of systematical errors. For the
implementation of the systematical errors and an explicit definition
of the $\chi^2$-function see appendix~\ref{app:syst}.  The calculations
have been performed with the {\sf GLoBES} software~\cite{Huber:2004ka,
Huber:2007ji, globesweb}, exploring the possibility of user-defined
$\chi^2$ in order to include the various systematics with the proper
correlations. A {\sf GLoBES} glb-file for our T2HK simulation with an
effective systematics treatment (see appendix~\ref{app:effectiveFD})
is available at~\cite{globesweb}.

\begin{figure}[p] \centering 
    \includegraphics[width=\textwidth]{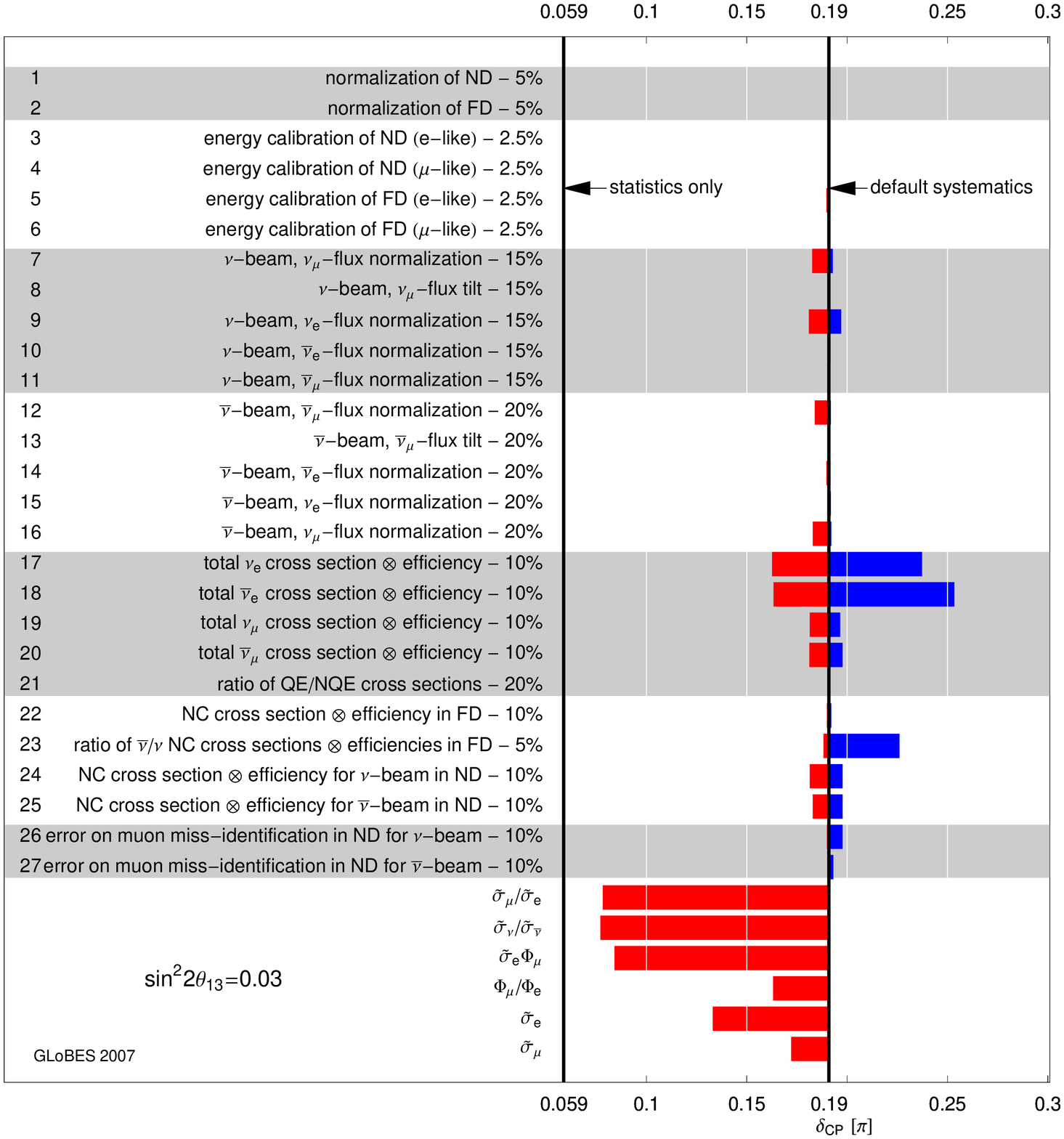}
    \mycaption{\label{fig:pulls} List of the systematical
      uncertainties and the adopted default values, as well as the
      impact of various systematics on the T2HK sensitivity to CPV for
      $\stheta = 0.03$. The abscissa shows the smallest $\delCP$ in
      $[0,\pi/2]$ for which CPV can be established at $3\sigma$. We
      show how the sensitivity is affected if each of the 27 pulls is
      switched off (red) or the error is multiplied by 5 (blue). The
      lower 6 rows show the impact when certain combinations of pulls
      are constrained at 2\%: the ratio of $\nu_e$ to $\nu_\mu$ cross
      sections times efficiencies (for neutrinos and anti-neutrinos),
      the ratio of neutrino and anti-neutrino cross sections times
      efficiencies (for $e$ and $\mu$-like events), the product of
      $\nu_\mu$ flux times $\nu_e$ cross section times $\nu_e$
      efficiency (for neutrinos and anti-neutrinos), the ratio of $e$
      to $\mu$ fluxes (for neutrinos and anti-neutrinos), $\nu_e$ and
      $\anu_e$ cross sections times efficiencies, $\nu_\mu$ and
      $\anu_\mu$ cross sections times efficiencies.}
\end{figure}

We include 27 uncorrelated errors listed in Fig.~\ref{fig:pulls}
together with the adopted default values. They include detector
normalisations and energy calibration errors, uncertainties on the
initial fluxes, cross section uncertainties which in our convention
include also uncertainties on the efficiencies, as well as errors on
NC backgrounds and muon miss-identification. A more detailed discussion
is given in appendix~\ref{app:syst} including also some motivations for
our default values. We stress that our defaults should by no means be
considered as the most realistic values, especially at the time when
the experiment is actually performed. Our assumptions are motivated by
the present situation (see appendix for references) or in other cases
are very conservative guesses. The purpose of our work is not to
advocate specific values for the systematics. Instead we want to
identify the crucial uncertainties which cannot be eliminated with the
help of the ND, and hence, for which solid external information is
required. 

\section{Results}
\label{sec:results}

\subsection{CP violation at T2HK}
\label{sec:CPV}

\begin{figure}[t] \centering 
    \includegraphics[width=0.6\textwidth]{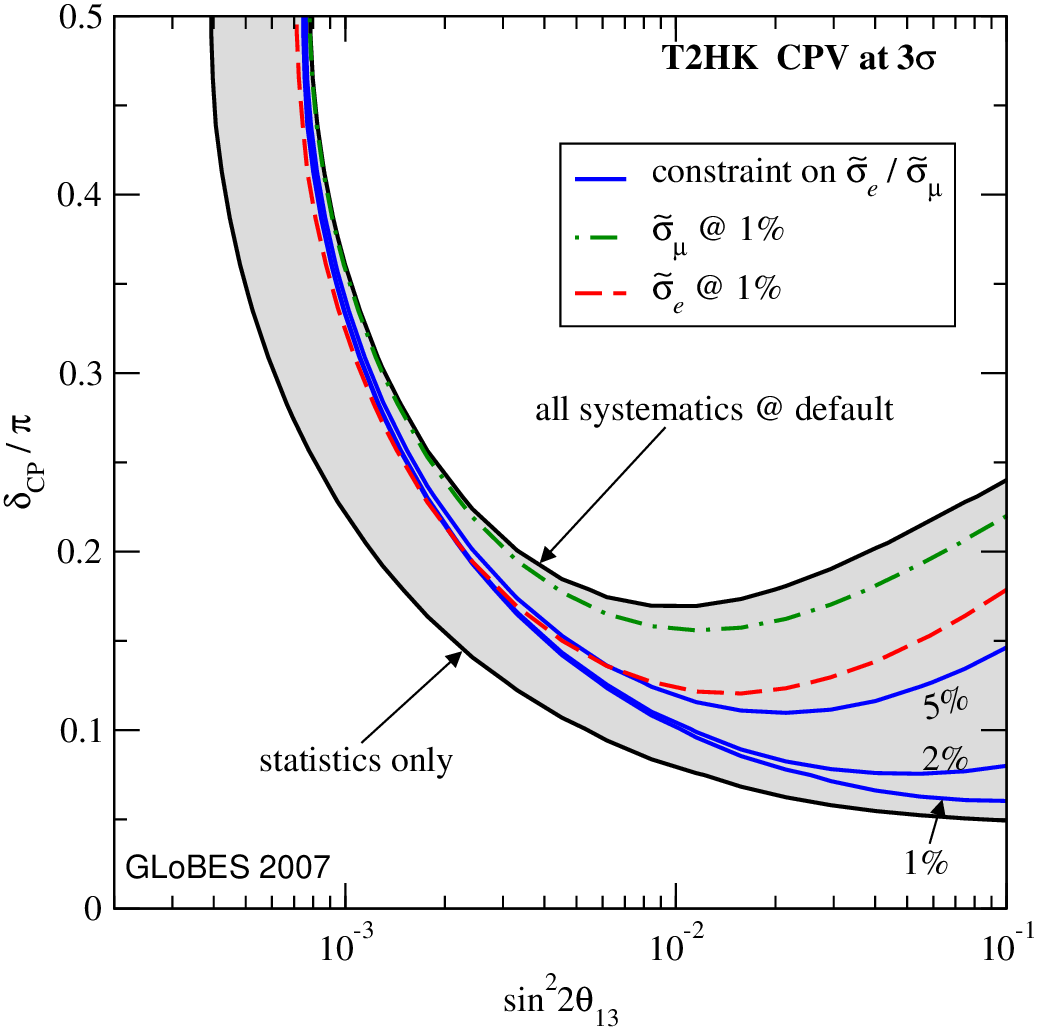} 
    \mycaption{\label{fig:T2HK-CPV} T2HK CPV sensitivity at $3\sigma$
    for our default choice of systematical errors according to
    Fig.~\ref{fig:pulls} and for statistical errors only (curves
    delimiting the shaded region).  We show also the sensitivity if
    certain constraints on the product of cross sections times
    efficiencies $\tilde\sigma$ are available: 1\% accuracies on
    $\tilde\sigma_{\nu_e}$ and $\tilde\sigma_{\nu_\mu}$ for neutrinos and
    anti-neutrinos, and 5\%, 2\%, 1\% accuracies on the ratios
    $\tilde\sigma_{\nu_e}/\tilde\sigma_{\nu_\mu}$ for neutrinos and
    anti-neutrinos.}
\end{figure}

In Fig.~\ref{fig:T2HK-CPV} we show the $3\sigma$ sensitivity of T2HK
for CPV. We restrict the analysis to the range $0 \le
\delCP^\mathrm{true} \le \pi/2$, and we neglect the sign($\Delta
m^2_{31}$) degeneracy. When calculating the $\chi^2$ for $\delCP = 0$
and $\pi$ we fix all oscillation parameters except from $\theta_{13}$
to their assumed true values $\Delta m^2_{31} = 2.4\times
10^{-3}$~eV$^2$, $\sin^2\theta_{23} = 0.5$, $\Delta m^2_{21} =
7.9\times 10^{-5}$~eV$^2$, $\sin^2\theta_{12} = 0.3$.
This allows to focus exclusively on the impact of systematics for the
discovery of CPV.\footnote{For the sensitivity to CPV as shown in
Fig.~\ref{fig:T2HK-CPV} the impact of the uncertainty on the
oscillation parameters (as well as on the matter density) is
negligible. See also the discussion later in connection with
Fig.~\ref{fig:regions}.}  The figure shows the sensitivity from
statistical errors only (lower black curve), which is obtained by
fixing all 27 systematic pulls to zero, as well as the sensitivity for
our default choice of systematical errors according to
Fig.~\ref{fig:pulls} (upper black curve). Clearly the impact is rather
dramatic for large $\theta_{13}$, in the region $\stheta \gtrsim
10^{-2}$. The effect of the various systematics in that region is
illustrated in Fig.~\ref{fig:pulls}, where we show what happens to the
smallest $\delCP$ for which CPV can be established at $3\sigma$ if
each single pull is switched off one by one (red bars) or multiplied
by a factor of 5 (blue bars), assuming $\sin^22\theta_{13}=0.03$.  No
single pull has a large impact on its own, which highlights the
importance of a comprehensive treatment of a large number of possible
error sources.

In Fig.~\ref{fig:T2HK-CPV} we show the case when rather precise
information (a hypothetical 1\% error) is available for either the
$\nu_\mu$ and $\anu_\mu$ effective cross sections or the $\nu_e$ and
$\anu_e$ effective cross sections.  Future cross section experiments
such as {\it e.g.}\ MINER$\nu$A~\cite{minerva} or
SciBooNE~\cite{AguilarArevalo:2006se} aim for a 5\% accuracy on the
absolute $\nu_\mu$ cross section. In section~\ref{sec:fluxes-ND} we
explore under which circumstances the near detector itself can perform
an accurate cross section measurement. Note, that the effective cross
section is defined as the product of cross section and
efficiency. Therefore, also the efficiencies would have to be known to
better than $1\%$.  Apparently only a marginal improvement is possible
for precisely known $\nu_\mu$ effective cross sections. This is not
very surprising since for $\nu_\mu$ the near detector can indeed
cancel a large fraction of the associated errors. Knowing the $\nu_e$
effective cross section would be helpful ({\it c.f.}\ also
Fig.~\ref{fig:pulls}), since here the near detector provides very
limited information, and the signal is directly proportional to this
cross section. However, clearly this information alone cannot resolve
the bulk of the systematics problem.

In contrast, the situation improves significantly if external
information is available on the ratio of the effective cross sections
$\tilde\sigma_R \equiv \tilde\sigma_{\nu_e}/\tilde\sigma_{\nu_\mu}$
for neutrinos and anti-neutrinos. Such information can come either
from theoretical calculations (for the cross section only, see
Sec.~\ref{sec:xsec}) or dedicated experiments. In order to perform
this analysis we take into account a correlation matrix between the
pulls corresponding to the effective cross sections, which imposes a
constraint on the ratio.  Hence, we replace the uncorrelated penalty
terms for the pulls in Eq.~(\ref{eq:chisq-final}) by a matrix
correlating the relevant pulls.  In Fig.~\ref{fig:T2HK-CPV} curves are
shown corresponding to an error on $\tilde\sigma_R$ of 5\%, 2\% and
1\%. Clearly, at the largest values of $\theta_{13}$ the error budget
is dominated by $\tilde\sigma_R$, and constraining this quantity
basically would allow to recover most of the statistical accuracy of
the experiment. This is in agreement with the discussion in
Sec.~\ref{sec:qualitative}, and can be understood by the fact the
relative size of CP effects is smallest for large $\theta_{13}$ and
hence the absolute accuracy of the prediction of the number of
oscillated $\nu_e$ events becomes very important. Any CP effect has to
be uncovered from this number or, more precisely, from its error. The
contribution of backgrounds to the number of $\nu_e$ events is small
and the contribution to the error is even smaller. The effect of a
constraint on $\tilde\sigma_R$ is also shown in the lower part of
Fig.~\ref{fig:pulls}, where we show the impact of constraining certain
combinations of systematics. There we display also the impact of a
constraint on the ratio of effective neutrino and anti-neutrino cross
sections, which has a very similar effect as a constraint on the
flavour ratio (even slightly more effective) for restoring the
statistics-only sensitivity, in agreement with the discussion related
to Eq.~(\ref{eq:combinations}).

\begin{figure}[t] \centering 
    \includegraphics[width=0.7\textwidth]{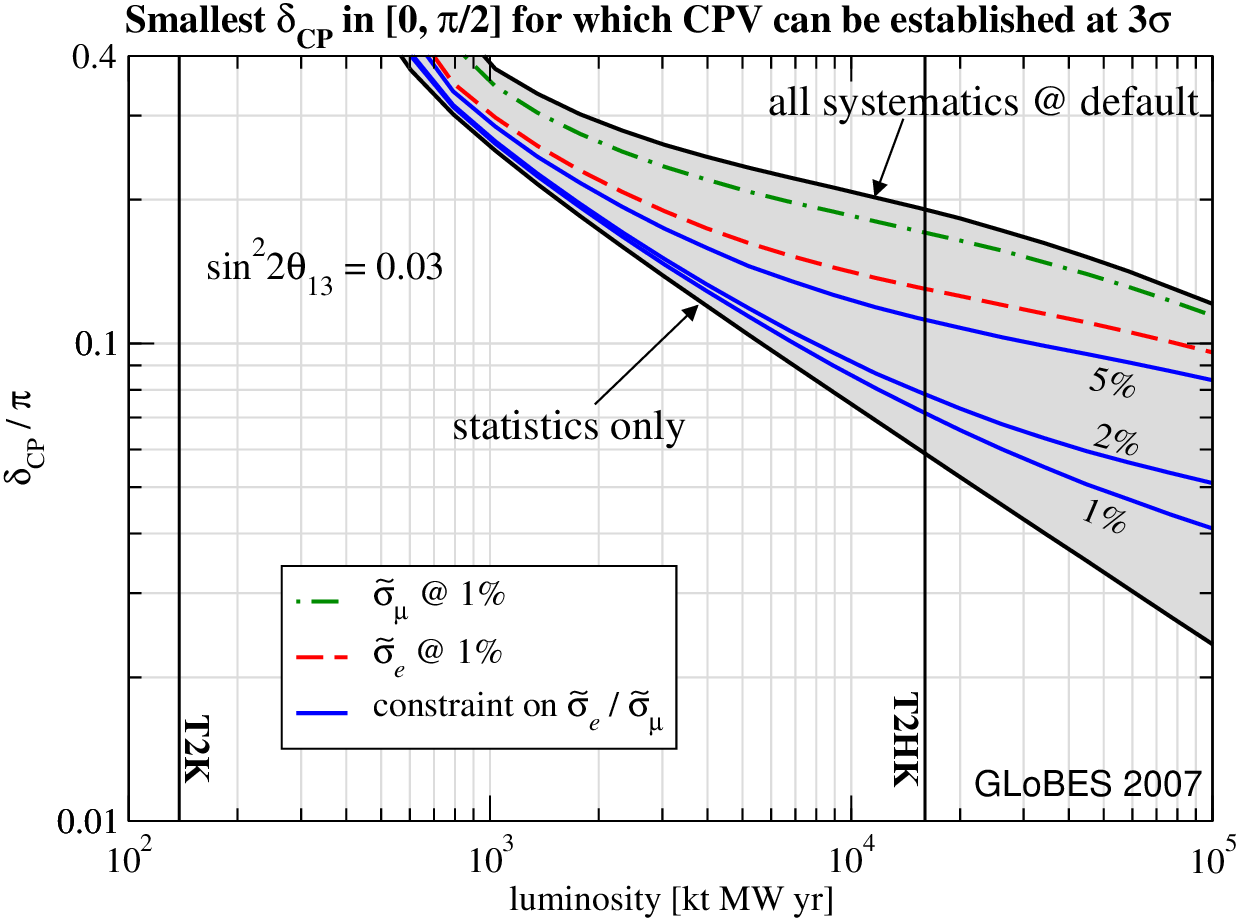}
    \mycaption{\label{fig:lumi-CPV} CPV sensitivity at $3\sigma$ as a
      function of exposure for a true value $\stheta = 0.03$ for our
      default choice of systematical errors according to
      Fig.~\ref{fig:pulls} and for statistical errors only (curves
      delimiting the shaded region). The ratio of neutrino to
      anti-neutrino running is kept constant at $1:3$.  Furthermore we
      show the sensitivity if certain constraints on the product of
      cross sections times efficiencies $\tilde\sigma$ are available:
      1\% accuracies on $\tilde\sigma_{\nu_e}$ and
      $\tilde\sigma_{\nu_\mu}$ for neutrinos and anti-neutrinos, and
      5\%, 2\%, 1\% accuracies on the ratios
      $\tilde\sigma_{\nu_e}/\tilde\sigma_{\nu_\mu}$ for neutrinos and
      anti-neutrinos.}
\end{figure}

Fig.~\ref{fig:lumi-CPV} shows the luminosity scaling of the
sensitivity to CPV in the region of large $\theta_{13}$ ($\stheta =
0.03$) with the same set of curves as in Fig.~\ref{fig:T2HK-CPV}.  For
this analysis we scale simultaneously the beam power and the FD mass
between the T2K (0.77~MW, 22.5~kt) and T2HK (4~MW, 500~kt) benchmarks
defined in Sec.~\ref{sec:simulation}, and we always assume a neutrino
(anti-neutrino) running time of 2~(6)~years, and hence a factor 8~yr
is included in the horizontal axis in Fig.~\ref{fig:lumi-CPV}.  The ND
mass is fixed at 0.1~kt.
Obviously, the impact of systematics becomes larger as the luminosity
increases.  Increasing luminosity is a possibility to compensate for
systematical errors, though a very costly one. For example, if an
external constraint on the ratio
$\tilde\sigma_{\nu_e}/\tilde\sigma_{\nu_\mu}$ at 2\% were available
T2HK could have 1/10 of the luminosity and still achieve the same
sensitivity to CPV as in the default configuration.  Thus, any
optimised experimental strategy has to find the right balance between
spending on measures to mitigate systematical effects and on
maximising the total luminosity. This may require a dedicated effort
since some of the necessary measurements may be external to the actual
oscillation experiment.

Let us briefly comment on the region of small $\theta_{13}$ close to
the sensitivity limit. Fig.~\ref{fig:T2HK-CPV} shows that in this case
a constraint on $\tilde\sigma_{\nu_e}/\tilde\sigma_{\nu_\mu}$ cannot
reduce the effect of systematics. The reason is that in this region
the precision on the background determines the sensitivity, see the
discussion in Sec.~\ref{sec:qualitative}. This leads to rather
different requirements than in the region of large $\theta_{13}$. The
relevant question is here to which accuracy the ND data can be used to
predict the background in the FD, {\it c.f.}\ Eq.~(\ref{eq:app-bg}).
This ability is limited by backgrounds in the ND ({\it e.g.,} NC, muon
miss-identification), as well as by statistics in the ND.

\begin{figure}[t] \centering 
    \includegraphics[width=0.47\textwidth]{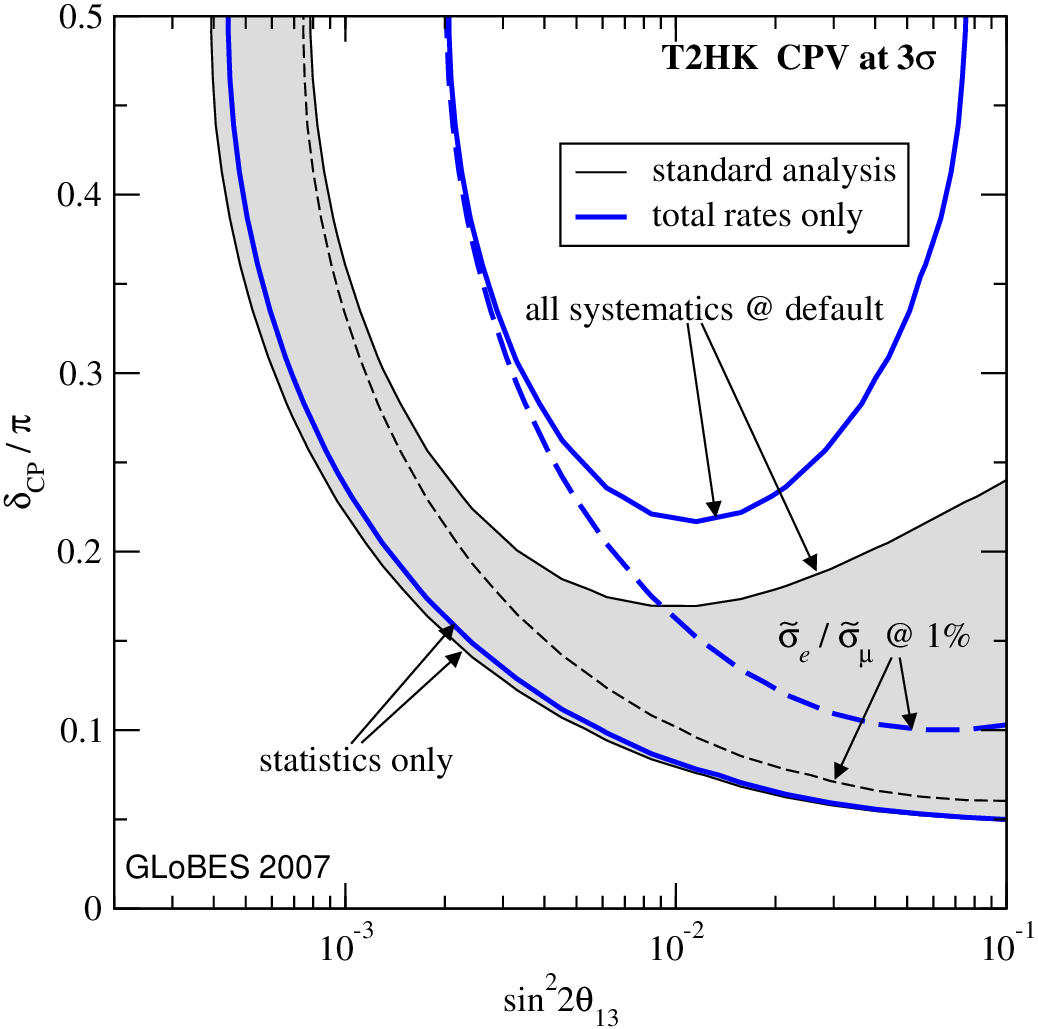} \quad
    \includegraphics[width=0.47\textwidth]{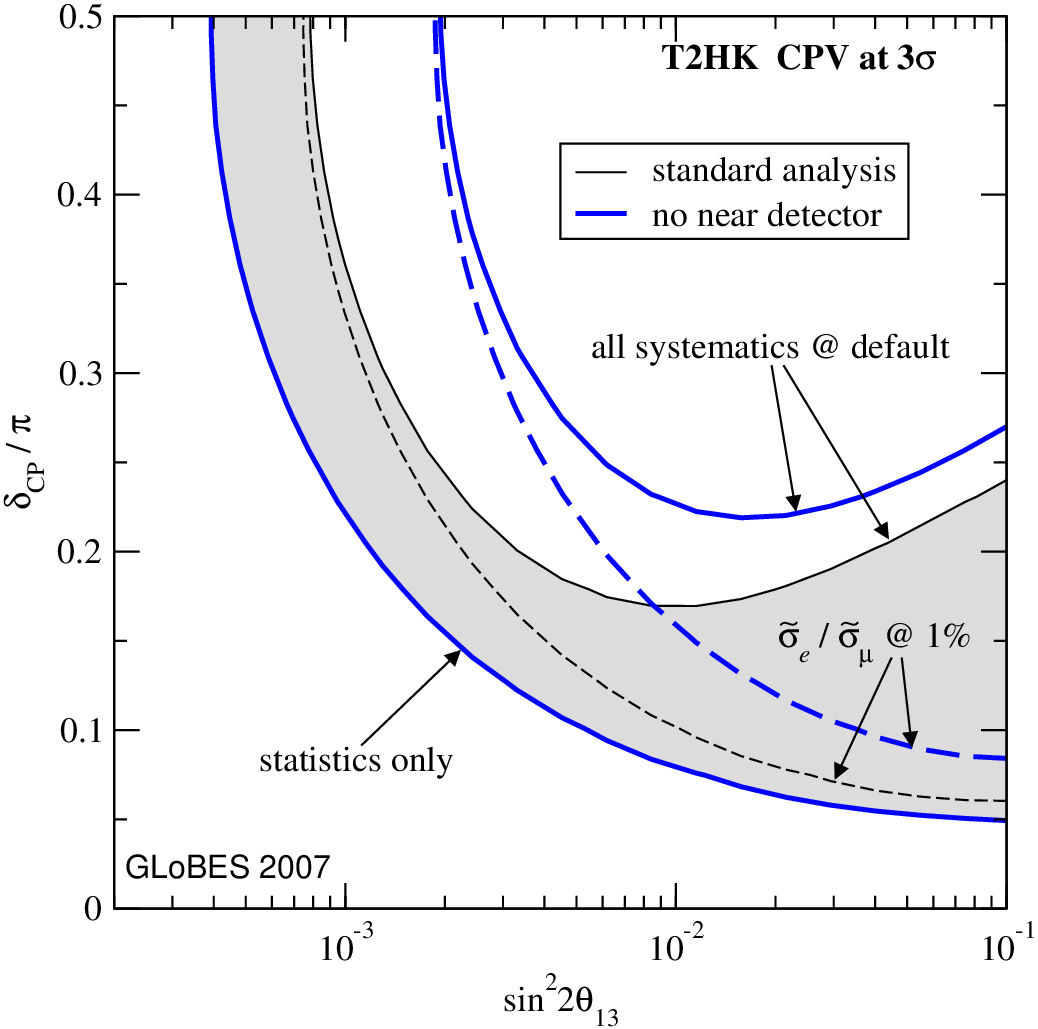} 
    \mycaption{\label{fig:T2HK-rate-ND} T2HK CPV sensitivity at
      $3\sigma$ for a total rate measurement only (left) and without a
      near detector (right) for our default choice of systematical
      errors according to Fig.~\ref{fig:pulls} and for statistical
      errors only. The dashed curves correspond to an external
      accuracy of 1\% on the ratios
      $\tilde\sigma_{\nu_e}/\tilde\sigma_{\nu_\mu}$ for neutrinos and
      anti-neutrinos. The shaded regions correspond to our standard
      analysis and are identical to the one in
      Fig.~\ref{fig:T2HK-CPV}.}
\end{figure}

Fig.~\ref{fig:T2HK-rate-ND} (left) shows the impact of spectral
information. The first observation is that the pure statistics
sensitivity hardly changes if only rate information is used, which
shows that the main oscillation physics is captured just in the total
number of events.\footnote{Note that spectral information is crucial
for resolving the so-called intrinsic degeneracy, see {\it
e.g.}~\cite{Huber:2002mx,Ishitsuka:2005qi,Campagne:2006yx}. However,
since the intrinsic degeneracy does not confuse CP violating and
conserving values of $\delCP$ it does not affect the sensitivity
to CP violation shown in Fig.~\ref{fig:T2HK-rate-ND}.} However, the
spectrum is important to disentangle oscillation effects from
systematics, especially in the regions of very small $\theta_{13}$ (to
measure the background in the ND) and very large $\theta_{13}$ (to
avoid confusion of the CPV signal with systematics).  This result
indicates that the systematics question in the context of a wide band
beam~\cite{Barger:2006vy,Diwan:2006qf,Barger:2007yw} might be
different, and it would be desirable to have a similar analysis also
for such a facility.

In Fig.~\ref{fig:T2HK-rate-ND} (right) we show the sensitivity to CPV
without any near detector. This plot highlights the importance of the
ND for small $\theta_{13}$, where it is needed to constrain the
background. However, the impact of the ND is somewhat smaller for
large $\theta_{13}$, $\stheta \gtrsim 0.05$, since here the question
of backgrounds is less important, whereas the main uncertainty comes
from the combinations given in Eq.~(\ref{eq:combinations}), {\it
  e.g.}\ the ratio $\tilde\sigma_{\nu_e}/\tilde\sigma_{\nu_\mu}$,
for which the ND provides only rather poor constraints.

\subsection{Constraints on neutrino fluxes and properties of the near detector}
\label{sec:fluxes-ND}

Let us now discuss the impact of some external knowledge on the
initial neutrino fluxes. In this case some of our default values might
appear slightly too conservative. Information on the fluxes requires
careful instrumentation of the beam line and data from dedicated
Hadron production experiments such as MIPP~\cite{Raja:2005sh} in the
case of MINOS, HARP~\cite{Catanesi:2007ig, harp:2007gt} for K2K and
MiniBooNE, or NA61/SHINE~\cite{Laszlo:2007ib} for T2K. For example,
in MINOS the goal is to constrain $\Phi_{\nu_\mu}$ at the 5\% level
using MIPP data. It is beyond the scope of this work to do a detailed
study of how well neutrino fluxes can ultimately be determined.  Here
we investigate the case of perfectly well known fluxes and how useful
that would be for the CPV measurement in T2HK.

\begin{figure}[t] \centering 
    \includegraphics[width=0.6\textwidth]{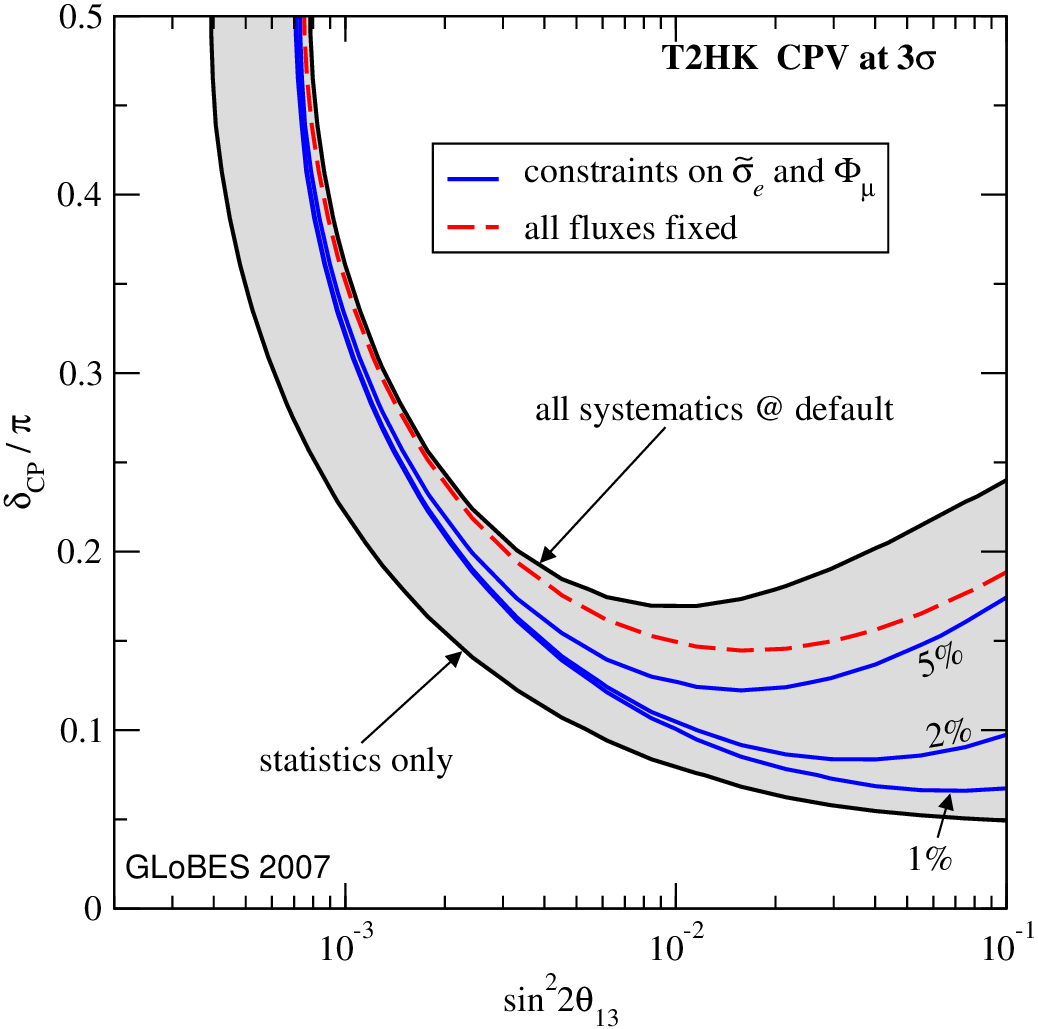}
    \mycaption{\label{fig:T2HK-flux} T2HK CPV sensitivity at
    $3\sigma$. We shows the impact of perfectly known fluxes, as well
    as constraints at 5\%, 2\%, 1\% on $\tilde\sigma_{\nu_e}$ and $\nu_\mu$
    fluxes, both for neutrinos and anti-neutrinos. The shaded region
    corresponds to our standard analysis and is identical to the one
    in Fig.~\ref{fig:T2HK-CPV}.}
\end{figure}

The dashed curve in Fig.~\ref{fig:T2HK-flux} corresponds to the case
of perfectly known fluxes (including all sub-dominant flavours in each
beam) with all other systematics at the default values. The
sensitivity improves slightly for $\stheta \gtrsim 0.01$, but clearly
this information is not enough to significantly address the
systematics problem. The reason can again be understood from the
discussion in Sec.~\ref{sec:qualitative}. Eq.~(\ref{eq:app-sig}) shows
that the uncertainty on $\tilde\sigma_{\nu_e} /
\tilde\sigma_{\nu_\mu}$ remains, irrespectively of the uncertainty on
the fluxes. As mentioned in the paragraph after
Eq.~(\ref{eq:combinations}) flux information is only useful in
combination with a constraint on the effective $\nu_e$ cross section.
This is confirmed by the blue curves in Fig.~\ref{fig:T2HK-flux},
which show that the impact of systematics can efficiently be reduced
by accurate external information on both, $\Phi_{\nu_\mu}$ and
$\tilde\sigma_{\nu_e}$ (for neutrinos and anti-neutrinos). Note that
this information is not provided by the ND, but has to come from
sources outside the considered setup. This seems especially difficult
for $\nu_e$ and $\anu_e$ cross sections.

\begin{figure}[t] \centering 
    \includegraphics[width=0.7\textwidth]{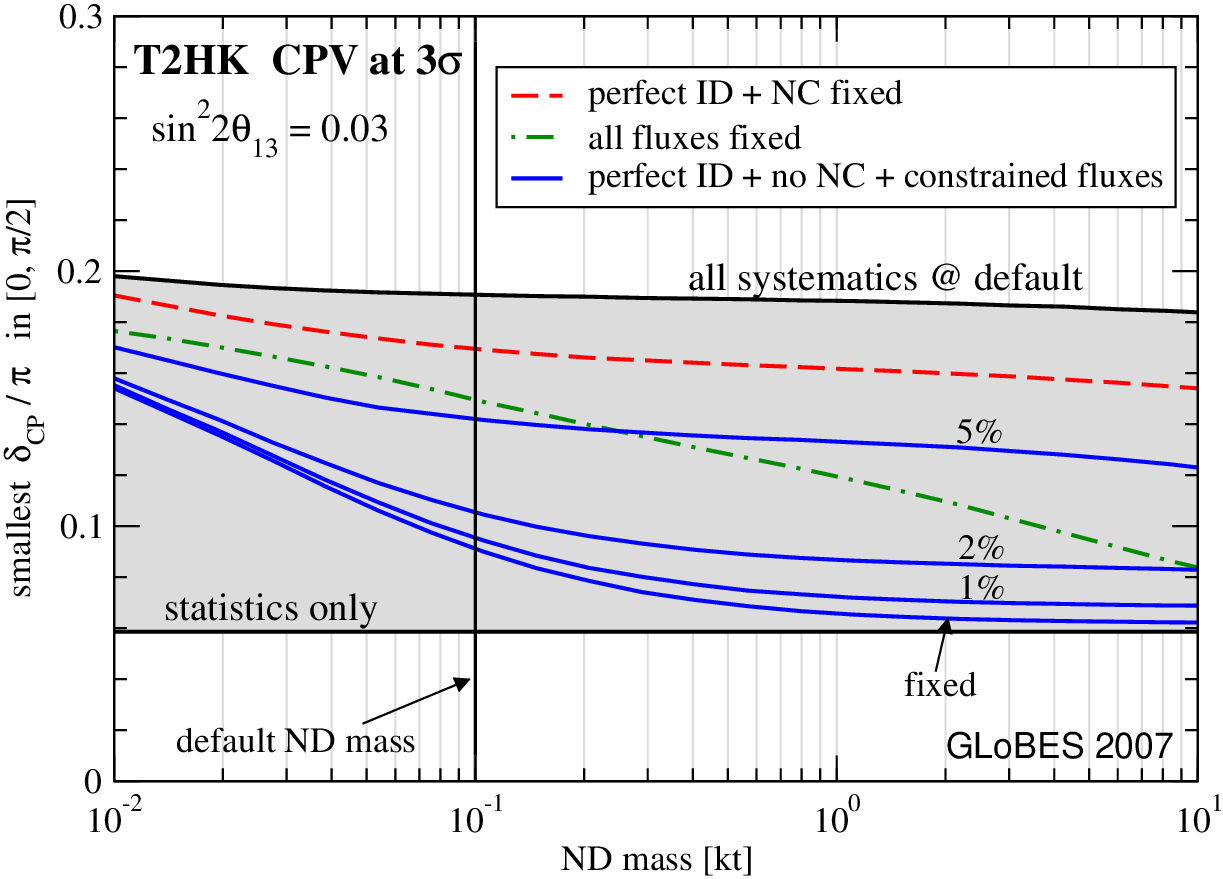}
    \mycaption{\label{fig:NDmass} CPV sensitivity at $3\sigma$ for
    T2HK as a function of the near detector mass for a true value
    $\stheta = 0.03$ for our default choice of systematical errors
    according to Fig.~\ref{fig:pulls} and for statistical errors only
    (curves delimiting the shaded region). The red/dashed curve
    corresponds to a ND with perfect $e/\mu$ separation and with
    perfectly known NC background, but all other systematics at
    default. The green/dash-dotted curve corresponds to the
    standard ND but we assume that all fluxes are perfectly known. For
    the blue/solid curves we assume a ND with perfect $e/\mu$
    separation and without any NC background plus some knowledge on
    the fluxes according to the labels.}
\end{figure}

Let us elaborate more on the somewhat surprising result, that knowing
the fluxes has such a small impact. Indeed, in this case one may
expect that the ND should be able to provide a measurement of
$\tilde\sigma_{\nu_e}$ via the intrinsic $\nu_e$ component in the beam
(which is assumed to be known perfectly). This argument is true in
principle, however, we find that within our implementation the NC
background and the muon miss-identification in the ND plus statistical
errors in the ND are enough to spoil this measurement. We illustrate
this effect in Fig.~\ref{fig:NDmass} by showing the CPV sensitivity
for $\stheta = 0.03$ as a function of the ND mass, while keeping our
T2HK default values for beam power, FD mass, and $\nu/\anu$ running
times constant. First, we note that for all systematics at default the
sensitivity depends very little on the size of the ND, which is
consistent with Fig.~\ref{fig:T2HK-rate-ND} (right). Second, the curve
for known fluxes (green/dash-dotted) shows a modest gain in
sensitivity at the default ND mass of 0.1~kt from
Fig.~\ref{fig:T2HK-flux}, and some improvement with increasing the
mass. However, the situation clearly improves if a ``perfect'' ND
without muon miss-identification and NC background is assumed. In this
case it only depends on the statistical errors in the ND and on the a
priori accuracy of the fluxes, how well $\tilde\sigma_{\nu_e}$ can be
determined by the ND. This can be seen from the blue/solid curves in
Fig.~\ref{fig:NDmass}, corresponding to a ``perfect'' ND plus a
constraint on the fluxes, where the accuracy indicated in the figure
is implemented as an uncorrelated error on each of the flux
components. From the plot we find that for a flux uncertainty of 1\%
the pure statistics sensitivity is nearly reached for ND masses of
about 1~kt. Let us add that such a ``perfect'' ND would also improve
the sensitivity to CPV at small $\theta_{13}$, since it would provide
an accurate determination of the background.

To summarise this discussion, if precise information on fluxes is
available (including the $\nu_e$ and $\anu_e$ components) it is
possible, in principle, to measure the electron cross sections in the
ND. This would lead to a strong reduction of the systematics impact,
since then both $\Phi_{\nu_\mu}$ and $\tilde\sigma_{\nu_e}$ were
known, which is one of the ``magic'' combinations identified in
Eq.~(\ref{eq:combinations}). To achieve this situation, in addition to
the flux information a careful design of the ND in terms of background
rejection and its size is necessary. In this respect we mention that
for T2K a liquid Argon detector with a fiducial mass of
$0.1\,\mathrm{kt}$ is foreseen at 2~km, which would allow to collect a
fairly clean sample of $\nu_e$ CC events~\cite{2kmLOI}.

\subsection{Determination of $\theta_{13}$ at T2K and T2HK}
\label{sec:th13}

\begin{figure}[t] \centering 
    \includegraphics[width=0.7\textwidth]{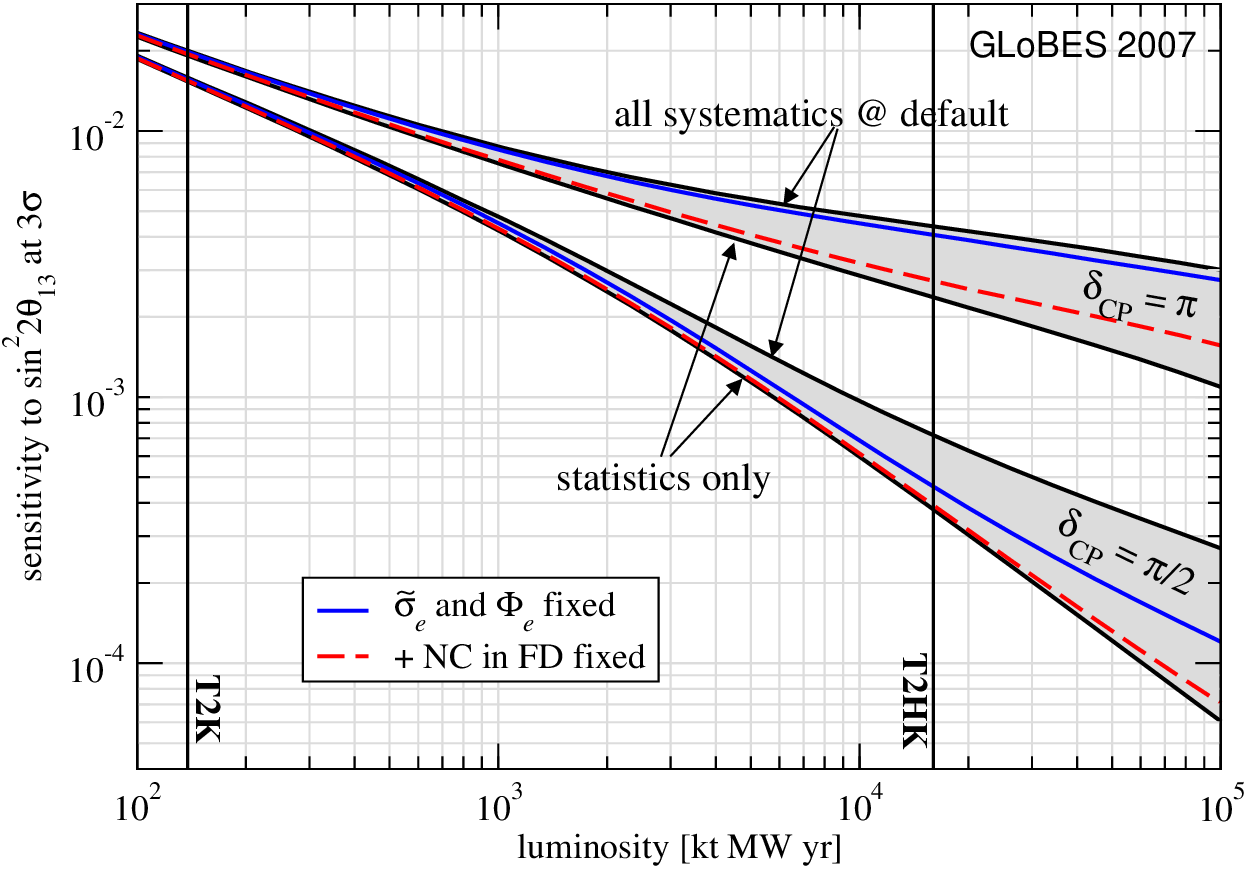}
    \mycaption{\label{fig:lumi-th13} Sensitivity at $3\sigma$ to a
      non-zero $\theta_{13}$ as a function of exposure for $\delCP =
      \pi/2$ and $\pi$ for our default choice of systematical errors
      according to Fig.~\ref{fig:pulls} and for statistical errors
      only (curves delimiting the shaded region). The ratio of
      neutrino to anti-neutrino running is kept constant at $1:3$.
      Furthermore, we show the sensitivity obtained without
      uncertainty on the intrinsic beam background (by fixing
      $\tilde\sigma_{\nu_e}$ and the $e$-like fluxes) and without an
      uncertainty on the NC background in the far detector.}
\end{figure}

Let us now discuss the impact of systematics on the $\theta_{13}$
measurement. In Fig.~\ref{fig:lumi-th13} we show the smallest value of
$\stheta$ which can be distinguished from $\theta_{13} = 0$ as a
function of the luminosity, assuming two representative values for the
CP phase which correspond roughly to the best and worst sensitivity.
The first observation is that for T2K phase~I systematics have only a
small impact, since this measurement is largely dominated by
statistics. Numerically we find that the sensitivity of T2K decreases
from $\stheta = 0.0167$ to 0.0172 for $\delCP = \pi/2$, and from
$\stheta = 0.0206$ to 0.0214 for $\delCP = \pi$.
For T2HK systematics have a non-negligible impact on the $\theta_{13}$
discovery reach.  The situation is very similar to CPV for small
$\theta_{13}$, and the corresponding discussion in Sec.~\ref{sec:CPV}
largely applies also for the $\theta_{13}$ discovery: for this
measurement the background dominates in Eq.~(\ref{eq:FDe}), and hence
the uncertainty on the background is the most relevant
systematics. Its impact is controlled by the ability of the ND to
predict the background in the FD. We show in Fig.~\ref{fig:lumi-th13}
also curves assuming a perfectly known $\nu_e$ beam background, and no
uncertainty at all on the background ({\it i.e.}, fixing the $\nu_e$
beam contamination as well as the NC background). If the total
background is fixed the sensitivity is close to the pure statistics
case. It is interesting to note that for the two examples of $\delCP$
shown in the figure the importance of beam and NC backgrounds is
different. This is an effect of the spectral shapes of the signal
relative to the background, since the spectrum of the signal depends
on the value of $\delCP$, and also beam and NC backgrounds have rather
different shapes.

\begin{figure}[t] \centering 
    \includegraphics[height=7cm]{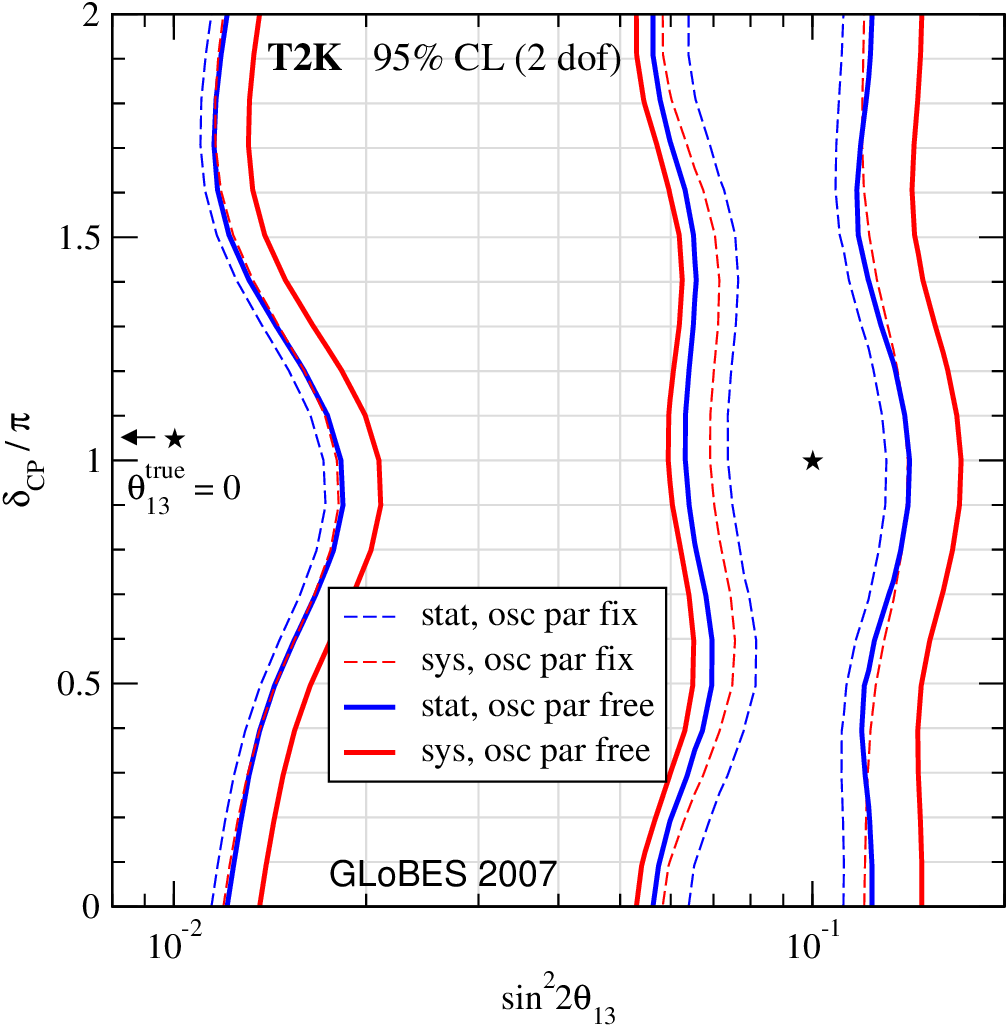} \quad
    \includegraphics[height=7cm]{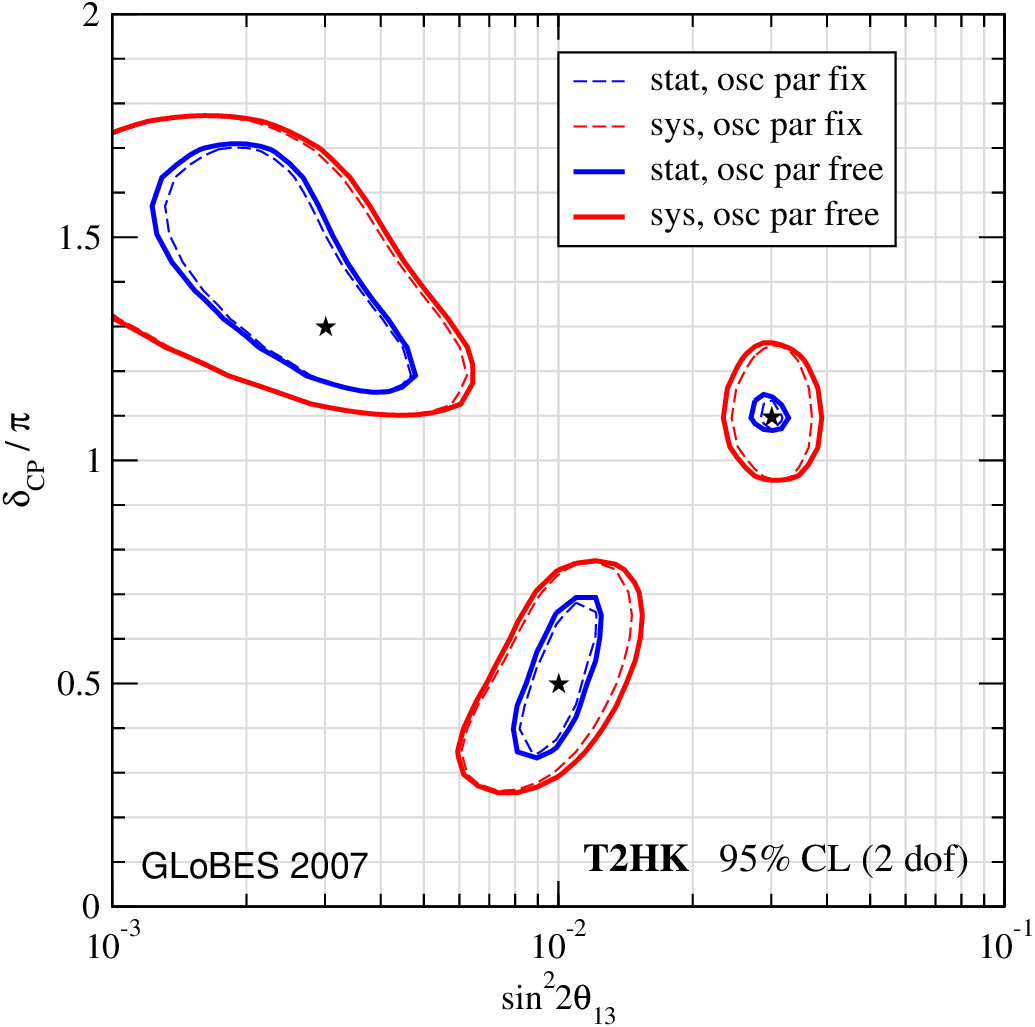} 
    \mycaption{\label{fig:regions} Allowed regions in the plane of
    $\stheta$ and $\delCP$ for T2K (left) and T2HK (right) for some
    example choices for the input values marked by stars in the
    figures. We show the allowed regions for all combinations of
    statistical errors only, systematics according to
    Fig.~\ref{fig:pulls}, all other oscillation parameters fixed, and
    free (where for the solar parameters we impose present
    errors). For regions labels ``osc par free'' we allow also for a
    5\% uncertainty on the matter density, which however has
    a negligible impact on the results.  The sign($\Dmq_{31}$)
    degeneracy is neglected, and $\theta_{23}^\mathrm{true} = \pi/4$.}
\end{figure}

Fig.~\ref{fig:regions} shows the allowed region in the space of
$\stheta$ and $\delCP$ obtained by T2K (left) and T2HK (right) for
some example points of ``true'' parameter values. As expected, for T2K
the impact of systematics is small, though not negligible in this
case. Furthermore, we show that for T2K the uncertainty on the other
oscillation parameters has a sizable impact on the allowed region. We
have checked that this effect comes entirely from the atmospheric
parameters $\Dmq_{31}$ and $\theta_{23}$. Apparently the disappearance
channel does not provide enough accuracy on these parameters to avoid
an effect on the $\theta_{13}$ determination. For the solar parameters
the accuracy from present data is sufficient to eliminate any effect
on the results shown in Fig.~\ref{fig:regions}.

On the other hand, for T2HK the impact of the correlations with the
other oscillation parameters is negligible, since the high statistics
sample of the disappearance channel pins down the atmospheric
parameters with high precision. In contrast systematics are much more
important. For example, for our test point at large $\theta_{13}$
($\stheta = 0.03$ and $\delCP = 1.1\pi$) the errors on $\stheta$ and
$\delCP$ are roughly a factor three larger if systematics are included.
Clearly in this case the inclusion of systematics would spoil the CPV
discovery.

\subsection{T2KK}
\label{sec:t2kk}

We have tested also the case when part of the HK detector is moved to
Korea, at a baseline of 1050~km. For this analysis we assume that the
second FD is located at the same off-axis angle as the first one, like
in~\cite{Ishitsuka:2005qi,Kajita:2006bt}. For our standard scenario
(``T2KK'') we assume a 250~kt detector, both in Kamioka and Korea.
The $\chi^2$ construction for the second FD is completely analogous to
the ones for the first FD, and we take into account the proper
correlations of systematics in the three detector system of ND, FD1,
FD2. We focus here on the CPV discovery in order to compare the T2KK
and T2HK performances. Needless to say, that a main motivation for
having a detector in Korea is to measure the neutrino mass hierarchy
which we do not discuss here. The hierarchy determination in turn
reduces the impact of degeneracies, which we have not included here in
order to focus on the impact of systematics. Such considerations have
to be taken into account when evaluating the overall potential of T2KK
(which is not the purpose of our discussion).

\begin{figure}[t] \centering 
   \includegraphics[width=0.9\textwidth]{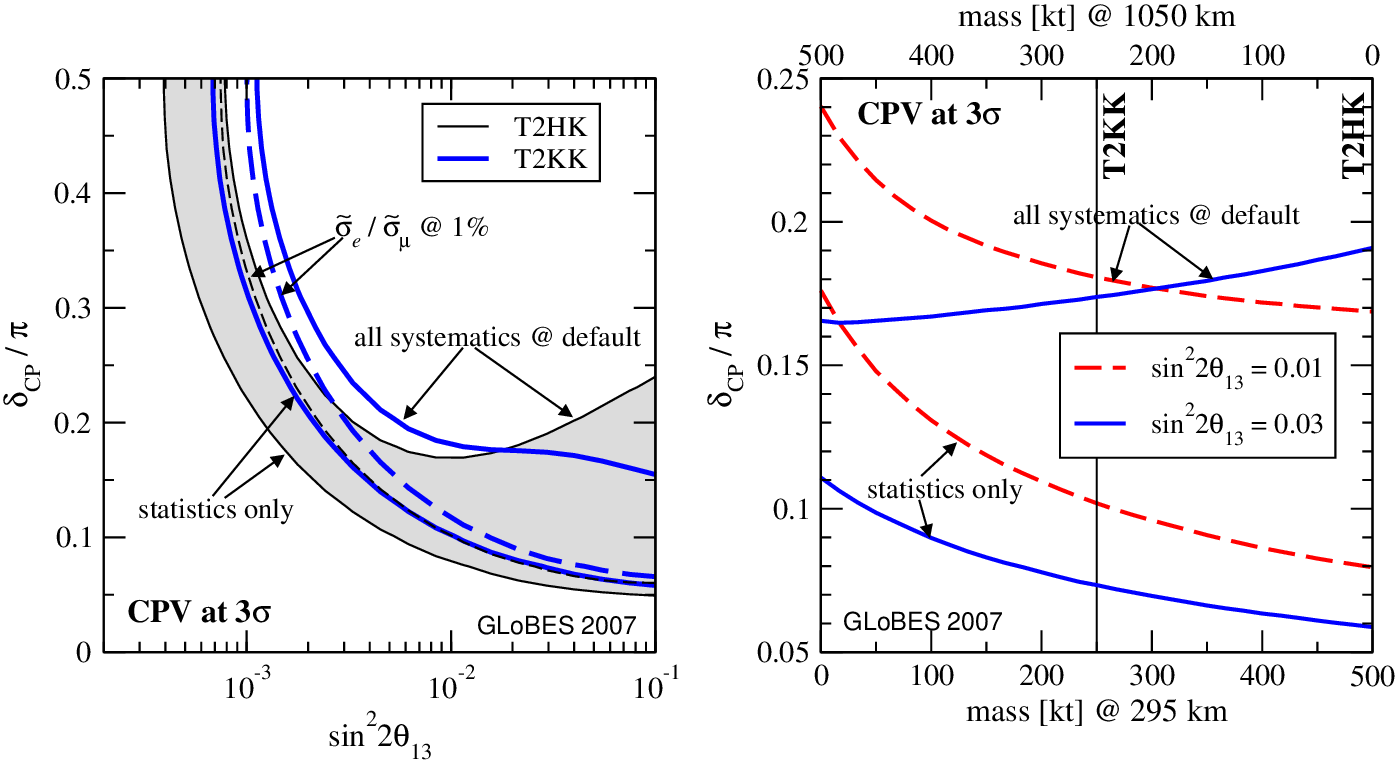}
    \mycaption{\label{fig:T2KK-CPV} Left hand panel: sensitivity to
    CPV at $3\sigma$ for T2HK and T2KK.  Right hand panel: sensitivity
    to CPV at $3\sigma$ for two values of $\stheta$ by changing the
    detector mass between Kamioka and Korea.}
\end{figure}

In the left hand panel of Fig.~\ref{fig:T2KK-CPV} the CPV discovery
reach of T2KK (blue lines) is shown in comparison to T2HK (shaded
region). We find that the pure statistics sensitivity is slightly
worse for T2KK. If all systematics are put at our default values
splitting the detector yields a somewhat better robustness with
respect to systematical uncertainties at large
$\sin^22\theta_{13}>10^{-2}$. If precise information on
$\tilde\sigma_{\nu_e}/\tilde\sigma_{\nu_\mu}$ is available T2HK and
T2KK perform rather similar.
We do not observe a particular cancellation of systematics beyond that
already present between near and far detector. Having two baselines
makes the physics signal more distinct and unique and hence it is
harder for systematical effects to mimic it for large $\theta_{13}$. A
similar result was found with a simplified analysis
in~\cite{Barger:2007jq}. Based on these results, it seems not necessary
to demand that the two detectors are identical or are located at the
same off-axis angle. This is especially important in the context of
the results in~\cite{Hagiwara:2005pe,Hagiwara:2006vn}, which indicate
that a more on-axis location for the detector in Korea would greatly
enhance the sensitivity to the mass hierarchy. The larger background
present at a more on-axis location maybe tolerable, especially if
improved algorithms for $\pi^0$ identification are used. This issue
has been extensively studied in the context of a wide band beam in the
US~\cite{Barger:2006vy,Diwan:2006qf,Barger:2007yw}.
 
In the right hand panel of Fig.~\ref{fig:T2KK-CPV} we show how the
discovery reach changes for various distributions of
$500\,\mathrm{kt}$ fiducial mass between the Korea and Kamioka sites.
In the case of statistics only (and neglecting the impact of the
hierarchy degeneracy) the conclusion would be that CPV is best
discovered by putting all mass to Kamioka. This conclusion changes in
presence of systematical errors and now it depends on $\theta_{13}$
whether T2KK or T2HK performs better.

\section{Summary and discussion}
\label{sec:conclusions}

We have studied the impact of a large number of possible systematical
errors on the ability of a superbeam experiment to discover CP
violation. As a specific example we chose T2HK, however our main
results should be applicable to all superbeam experiments using a
narrow band beam. We implemented a realistic description of the far
detector and included for the first time a near detector in such a
study. The emphasis of this work is not to predict the actual
sensitivity of a given experiment nor the actual size of systematical
errors, but to show that under semi-realistic assumptions the effects
are large and need to be studied in more detail.  We find that the
cancellation of systematics between near and far detectors remains
incomplete for the appearance channel, due to a lack of information in
the near detector on the final state. 

In this respect we have identified two qualitatively different regimes
depending on the size of $\theta_{13}$. For small values, close to the
sensitivity limit, the main issue is the uncertainty on the
background. In this case the performance depends on the ability of the
near detector to predict the background in the far detector. In the
regime of large $\theta_{13}$ ($\stheta \gtrsim 0.01$, which is
probably the more interesting range for this type of experiments)
backgrounds are a minor issue and the uncertainty on the signal itself
dominates. We find that the impact of systematics even with a near
detector is rather strong in this regime. For instance, for T2HK at
$\sin^22\theta_{13}=0.1$ the smallest $\delCP$ for which CPV can be
established increases from $0.05\pi$ for the statistics only case to
$0.24\pi$ when systematics are included.

However, we were able to identify crucial combinations of
parameters, which, when well constrained (at the level of $\lesssim
2\%$) can restore the sensitivity nearly to its statistics only value,
namely
\begin{itemize} 
\item
the ratios of the effective $\nu_\mu$ and $\nu_e$ cross sections
$\tilde\sigma_{\nu_\mu}/\tilde\sigma_{\nu_e}$ for neutrinos and
anti-neutrinos, or
\item
the ratios of the effective cross sections between
neutrinos and anti-neutrinos, for $\nu_e$ and $\nu_\mu$, or
\item
the initial flux of $\nu_\mu$ and the effective $\nu_e$ cross section,
both for neutrinos and anti-neutrinos.
\end{itemize}
With the effective cross section $\tilde\sigma$ we mean here the
product between physical cross section and detection efficiency.  The
success of a superbeam experiment in the regime $\stheta \gtrsim 0.01$
will depend to a significant degree on the information
available on these combinations.

Theoretical cross section calculations indicate that the uncertainty
on the ratio $\sigma_{\nu_\mu}/\sigma_{\nu_e}$ might actually be at
the level of few percent in the T2K energy range of around
700~MeV. However, this result has not been tested experimentally. We
stress that this would be a crucial input in the analysis of a
superbeam experiment which is not based on any data. Future cross
section experiments such as for example MINER$\nu$A may provide a
measurement of $\sigma_{\nu_\mu}$ at the 5\% level. However, from
present perspective it seems difficult to obtain a precise measurement
for electron neutrino---and especially for electron anti-neutrino
cross sections, which are essential for predicting the appearance
signal. Maybe the only places where these cross sections can be
measured in the relevant energy range are beta beams or a neutrino
factory. Note that the absolute normalisation of the cross sections is
needed, which always is subject to uncertainties on initial
fluxes. Precise information on fluxes may be obtained from Hadron
production experiments, such as MIPP, HARP or NA61/SHINE.

Apart from CP violation in T2HK, we find that systematics play a minor
role for the $\theta_{13}$ discovery sensitivity in T2K (phase~I),
since this measurement is dominated by statistical errors.
For the T2KK setup, where half of the HK detector is moved to Korea,
we find that the second far detector helps somewhat in reducing the
effect of systematics on the CP violation sensitivity at large
$\theta_{13}$. However, this effect does not come from a cancellation
of systematics, but from a more robust oscillation signal in the
very-far detector. Hence it seems not necessary to demand that the two
far detectors are identical or placed at the same off-axis angle.

In order to focus on the impact of systemtaics we have neglected the
hierarchy degeneracy in our study. It is known that for T2HK the
presence of the degenerate solution reduces the sensitivity to CPV in
a certain range of the parameter space. We have checked that our
conclusion on systematics is not changed due to the degeneracy, since
also the fake solution is affected in a similar way by systematics as
the true one. We stress that for a full evaluation of the CPV
sensitivity and a comparison to other experimental options (such as
{\it e.g.}\ T2KK) degeneracies have to be included.

Before concluding we add here a few thoughts on whether and how our
results may be extrapolated for other facilities, beyond T2HK. Our
results indicate that spectral information plays an important role in
limiting the effect of systematical uncertainties. This suggests that
the behaviour of a wide band superbeam will be different. Without a
detailed simulation it is hard to estimate quantitatively whether the
impact of systematics is significantly less than in the case of the
off-axis configuration considered here, and clearly investigations
along these lines would be an interesting topic for future work.

For a beta beam in principle similar considerations apply as in the
case of the superbeam, however there are some important
differences. First, the initial flux of electron neutrinos is known to
good precision. Second, since the signal here is $\nu_\mu$ appearance,
the relevant cross sections are much easier to measure at a MINER$\nu$A
type experiment. Hence, it seems easier to constrain the beta beam
equivalent of the last combination of quantities listed above, namely
$\nu_e$ fluxes and $\nu_\mu$ cross sections.  As already mentioned, a
close detector at a beta beam would probably be an ideal place to
measure the electron cross sections needed for a superbeam experiment.

The optimal facility concerning systematics seems to be a neutrino
factory. In this case intense fluxes of all four flavours
$\Phi_{\nu_e}, \Phi_{\anu_e}, \Phi_{\nu_\mu}, \Phi_{\anu_\mu}$ are
available at the near detector, and they are known with very good
precision. Hence, all cross sections can be measured accurately at the
near detector, which allows to predict the appearance signal in the
far detector basically free of systematics on fluxes and cross
sections.\footnote{We note that in case of the neutrino factory
another important systematics (at large $\theta_{13}$) is the
uncertainty on the matter density. Its effect on the CP violation
sensitivity has been discussed in Ref.~\cite{Huber:2006wb} together
with possibilities to reduce it.} Nonetheless, it would be useful to
actually prove this by explicit calculation.

\begin{acknowledgments}
  We thank Michele Maltoni for discussions in the initial phase of
  this work and for providing the routine used for the minimisation of
  the pull $\chi^2$, Eligio Lisi for useful comments on a draft of
  this paper, Davide Meloni for providing the results
  of~\cite{Benhar:2006nr} in machine readable format, Constantinos
  Andreopoulos for providing us with GENIE~\cite{genie} results,
  Jocelyn Monroe for invaluable clarifications about the MiniBooNE
  $\nu_e$ analysis, Morgan Wascko for useful comments about flux
  uncertainties, Debbie Harris for a careful reading and a number of
  suggestions to improve this manuscript, and Natalie Jachowicz for
  most useful discussions on neutrino cross sections.  We acknowledge
  the support of the European Community-Research Infrastructure
  Activity under the FP6 ``Structuring the European Research Area"
  program (CARE, contract number RII3-CT-2003-506395).
\end{acknowledgments}

\appendix
\section{Experiment simulation and systematics treatment}
\label{app:simulation}

\subsection{Detector simulation}
\label{app:detector}

The $\nu_e$ and $\anu_e$ appearance signals in the far detector (FD)
are calculated in the following way. We take into account
quasi-elastic (QE) as well as non-quasi-elastic (NQE) charged current
events using cross sections from the NUANCE v3r503 event generator
\cite{Nuance}.  However, we require that only a single ring is visible
in the detector which strongly reduces the number of NQE events. At
the generator level the requirement is that just one particle momentum
is above the \v{C}erenkov production threshold. We take into account
an energy dependent efficiency for $e$-like
events~\cite{Itow:2001ee}. Since this efficiency is the product of
single ring events plus particle identification for $\nu_e$ events, we
disentangle the one ring efficiency computed with NUANCE, properly
weighted between QE and NQE events, and extract a particle
identification efficiency that we assume to be the same for QE, NQE,
$\nu_e$ and $\anu_e$ events. The absolute efficiency is normalised in
order to reproduce the total number of signal events provided in
Tab.~2 of Ref.~\cite{Itow:2001ee} for given oscillation parameters and
5~yr neutrino data in T2K. We assume the same efficiency function also
for anti-neutrino data. For $\mu$-like events we take a flat
efficiency of 0.9.

The following background sources are included for the $\nu_e$
appearance signal (and CP symmetric for the $\anu_e$ signal):
miss-identified neutral current (NC) events, the intrinsic $\nu_e$ and
$\anu_e$ beam contamination, $\anu_e$ events from oscillations of the
$\anu_\mu$ beam component, and a tiny background from miss-identified
muons from $\nu_\mu$ charged current (CC) events (at a rate of
0.1\%). For the NC background we extract the spectral shape and number
of events from Ref.~\cite{Itow:2001ee} by scaling to our
exposures. Lacking any detailed information for the anti-neutrino
mode, we assume the same size and spectrum for the NC background as in
the neutrino mode.

For the spectral analysis we take into account the energy
reconstruction for QE and NQE events via migration matrices calculated
by using NUANCE~\cite{Nuance} and SK reconstruction algorithms (see
Ref.~\cite{Maesaka:2005aj}, p.~139). We use 50 bins in true neutrino
energy from zero to 2~GeV mapped onto 8 bins in reconstructed neutrino
energy from 0.4 to 1.2~GeV. In total we apply 8 migration matrices:
for QE and NQE events for each neutrino flavour $\nu_e$, $\anu_e$,
$\nu_\mu$, $\anu_\mu$. Each matrix is normalised to take into account
the single ring efficiency. The migration is consistently applied to
signal and $\nu_e$ background events. Precise information on our FD
simulation including the migration matrices, backgrounds and
efficiencies can be recovered from the {\sf GLoBES} glb-file available
at~\cite{globesweb}.
 
For the near detector (ND) we assume the idealised situation that the
flux is identical to the one of the far detector for no oscillations
({\it i.e.}, perfect near-to-far extrapolation).  For definiteness we
take a 0.1~kt detector at a distance of 2~km, for T2K as well as for
T2HK. We use the same migration matrices and efficiencies as in the
far detector.  For the $\mu$-like events we assume no background
beyond the events from the $\nu_\mu$ and $\anu_\mu$ beam components.
For $e$-like events we take into account the beam intrinsic $\nu_e$
and $\anu_e$ fluxes, NC events, as well as miss-identified muons from
$\nu_\mu$ (or $\anu_\mu$) CC interactions with a rate of 0.1\%. For
the NC events we assume the same spectrum as in the FD with the
normalisation scaled according to the different mass and baseline of
the ND.

\subsection{$\chi^2$ definition and systematics}
\label{app:syst}

For the statistical analysis we adopt the following $\chi^2$ function based on
Poisson statistics in each bin:
\begin{equation}\label{eq:chisq-data}
  \chi^2_\mathrm{data} (\boldsymbol{\theta}, \xi_\alpha)
  = 
  2 \sum_{A = 1}^8 \sum_{i = 1}^8 
  \left[ T_i^A(\boldsymbol{\theta}, \xi_\alpha) - D_i^A + 
  D_i^A \ln \frac{D_i^A}{T_i^A(\boldsymbol{\theta}, \xi_\alpha)} \right] \,,
\end{equation}
where the index $A$ runs over the 8 data samples in our analysis
obtained by all combinations of FD/ND, $\nu/\anu$-beam, and
$e/\mu$-like events as given in Tab.~\ref{tab:samples}. The samples 1,
2 (3, 4) correspond to the appearance (disappearance) channels in the
FD, whereas samples 5--8 are the ND data.  For the T2KK analysis we
add 4 more data samples to the ones given in Tab.~\ref{tab:samples}
corresponding to the second FD.  In Eq.~(\ref{eq:chisq-data}),
$T_i^A(\boldsymbol{\theta}, \xi_\alpha)$ is the theoretical prediction
for energy bin $i$ in data sample $A$, depending on the oscillation
parameters $\boldsymbol{\theta}$ and the pulls $\xi_\alpha$
parameterising the systematic uncertainties. Taking only the leading
term in the (small) pulls, in general these predictions can be written
as\footnote{In our code we use a slightly more complicated pull
  dependence in order to make sure that the $T_i^A$ stay always
  positive, which, however, is equivalent to Eq.~(\ref{eq:theor}) at
  first order in $\xi_\alpha$.}
\begin{equation}\label{eq:theor}
  T_i^A(\boldsymbol{\theta}, \xi_\alpha) = 
  N_i^A(\boldsymbol{\theta}) + 
  \sum_\alpha \xi_\alpha \, \pi_{i\alpha}^A(\boldsymbol{\theta}) \,.
\end{equation}
Note that the $T_i^A$ for the ND ($A=5,\ldots,8$) do not depend on the
oscillation parameters, and hence, the corresponding terms in the
$\chi^2$ serve only to constrain the pulls $\xi_\alpha$. As usual,
the corresponding ``data'' $D_i^A$ is taken as the prediction $T_i^A$
at some assumed ``true values'' for the oscillation parameters,
$\boldsymbol{\theta}^\mathrm{true}$, and for zero pulls, {\it i.e.}, $D_i^A
= N_i^A(\boldsymbol{\theta}^\mathrm{true})$. The final $\chi^2$ is
obtained by adding the penalty terms for the pulls and minimising with
respect to them:
\begin{equation}\label{eq:chisq-final}
  \chi^2(\boldsymbol{\theta}) =
  \min_{\xi_\alpha} \left[
  \chi^2_\mathrm{data} (\boldsymbol{\theta}, \xi_\alpha) +
  \sum_{\alpha = 1}^{27} \left(\frac{\xi_\alpha}{\sigma_\alpha}\right)^2
  \right] \,.
\end{equation}
The pull minimisation is performed by using a routine developed by
Michele Maltoni.

\begin{table}
\begin{tabular}{c@{\quad}ccc@{\qquad\qquad}c@{\quad}ccc}
  \hline\hline
  $A$ & Detector & Beam & Flavour &
  $A$ & Detector & Beam & Flavour \\
  \hline
  1 & FD & $\nu$  & $e$-like   &  5 & ND & $\nu$  & $e$-like \\	 
  2 & FD & $\anu$ & $e$-like   &  6 & ND & $\anu$ & $e$-like \\	 
  3 & FD & $\nu$  & $\mu$-like &  7 & ND & $\nu$  & $\mu$-like \\
  4 & FD & $\anu$ & $\mu$-like &  8 & ND & $\anu$ & $\mu$-like \\
  \hline\hline
\end{tabular}
  \mycaption{\label{tab:samples} Data samples.}
\end{table}

In our analysis we include 27 independent pulls to account for
systematic uncertainties as listed in Fig.~\ref{fig:pulls}
together with the adopted default value for the errors
$\sigma_\alpha$. They are coupled to the theoretical predictions via
the couplings $\pi_{i\alpha}^A$ according to Eq.~(\ref{eq:theor})
accounting for the correct correlations. The pulls 1 and 2 describe
the normalisation uncertainties of FD and ND (fiducial mass),
correlated between $\nu$- and $\anu$-beams but uncorrelated between
the two detectors. Pulls 3--6 take into account the energy calibration
uncertainty, correlated between $\nu$- and $\anu$-beams but
uncorrelated between the two detectors and $e$- and $\mu$-like events.
Following Ref.~\cite{2kmLOI} we adopt a value of 2.5\%.
The pulls 7--16 account for the flux uncertainties, which are
correlated between the two detectors. For each beam we assume all four
flavour components to be uncorrelated. For the dominating flux we
include in addition a linear tilt on the spectral shape. These errors
are representative of the situation without a dedicated
hadron production experiment. They are about the same as K2K would
have had without the HARP data~\cite{Ahn:2006zz}.

Pulls 17--21 parametrise cross section uncertainties. We include an
uncorrelated normalisation uncertainty on the total CC cross section
for all four neutrino flavours $\nu_e$, $\nu_\mu$, $\anu_e$,
$\anu_\mu$.  Note, that pulls 17--21 also account
for the effect of the error on the efficiency. Since we assume
identical detectors for the FD and ND it seems justified to consider
the efficiencies correlated between the detectors, and hence we
consider the pulls 17--20 as the effective uncertainty including cross
section as well as efficiency uncertainties for the corresponding
event types. Pull 21 accounts for the uncertainty on the ratio of QE
and NQE cross sections, which we take fully correlated between
flavours and neutrino/anti-neutrino. K2K has measured this ratio in
different near detectors and has found a spread of 20\% among the
measurements~\cite{Ahn:2006zz}, which we adopt as our default value
for the uncertainty.

The uncertainty of NC events are taken to be completely uncorrelated
between ND and FD. For the FD we include pull 22 for the NC
normalisation correlated between $\nu$- and $\anu$-beams, whereas pull
23 accounts for the relative uncertainty. Pulls 24 and 25 account for
the NC backgrounds to $e$-like events in the ND, uncorrelated between
$\nu$- and $\anu$-beams. And finally, we include an uncertainty for
the rate of muon miss-identification in the ND, again uncorrelated
between $\nu$- and $\anu$-beams (pulls 26 and 27).
For the T2KK analysis we add 3 more pulls, one for the normalisation
and two for the energy calibration of the detector in Korea.

Let us give one explicit example, how the theoretical predictions
according to Eq.~(\ref{eq:theor}) are constructed, {\it e.g.}, for the
$\nu_e$ appearance channel in the FD ($A = 1$). In this case we have
\begin{equation}
  N_i^1 = n^{i,\mathrm{QE}}_{\nu_\mu\to\nu_e}
        + n^{i,\mathrm{NQE}}_{\nu_\mu\to\nu_e}
        + n^i_\mathrm{NC} 
        + n^i_{\nu_e-\mathrm{beam}} 
        + n^i_{\anu_e-\mathrm{beam}} 
        + n^i_{\anu_\mu\to\anu_e} 
        + n^i_\mathrm{miss-ID} \,.
\end{equation}
The $n^i_x$ correspond to the oscillation signal and the various
backgrounds as described above.  For simplicity we suppress the
dependence on the oscillation parameters, however, in the calculation
also oscillations of backgrounds are properly included. Then 
Eq.~(\ref{eq:theor}) reads
\begin{align}
  T_i^1  & = N_i^1 (1 + \xi_2) + 
             \xi_5 \, \pi^1_{i,\mathrm{calib\,FD},e} \nonumber\\
         & +\xi_7 \, (n^{i,\mathrm{QE}}_{\nu_\mu\to\nu_e}
                + n^{i,\mathrm{NQE}}_{\nu_\mu\to\nu_e}
                + n^i_\mathrm{NC} 
                + n^i_\mathrm{miss-ID}) 
           +\xi_8 \, \pi^1_{i,\mathrm{tilt}} \nonumber\\
         & +\xi_9 \, n^i_{\nu_e-\mathrm{beam}} 
           +\xi_{10} \, n^i_{\anu_e-\mathrm{beam}} 
           +\xi_{11} \, n^i_{\anu_\mu\to\anu_e}  \nonumber\\
         & +\xi_{17} \, (n^{i,\mathrm{QE}}_{\nu_\mu\to\nu_e}
                       + n^{i,\mathrm{NQE}}_{\nu_\mu\to\nu_e}
                       + n^i_{\nu_e-\mathrm{beam}} )
           +\xi_{18} \, (n^i_{\anu_e-\mathrm{beam}} 
                       + n^i_{\anu_\mu\to\anu_e})
           +\xi_{19} \, n^i_\mathrm{miss-ID} \nonumber\\
         & +\xi_{21} \, (n^{i,\mathrm{QE}}_{\nu_\mu\to\nu_e}
                       - n^{i,\mathrm{NQE}}_{\nu_\mu\to\nu_e})/2 \nonumber\\
         & +\xi_{22} \, n^i_\mathrm{NC} 
           +\xi_{23} \, n^i_\mathrm{NC}/2  \,.
\end{align}
Here $\pi^1_{i,\mathrm{calib\,FD},e}$ and $\pi^1_{i,\mathrm{tilt}}$
account for the energy calibration and $\nu_\mu$-flux tilt,
respectively. The 6 lines take into account normalisation and
calibration, $\nu_\mu$-flux uncertainty, uncertainties of the other
flux components, total cross section and efficiency uncertainties, the
QE/NQE ratio, and NC uncertainties. The last line for the $\anu$-beam
($A=2$) would read $+\xi_{22} \, n^i_\mathrm{NC} -\xi_{23} \,
n^i_\mathrm{NC}/2$ in order to account for the correlations of the NC
pulls as described above. The $T_i^A$ for the other samples are
defined in an analogous way. Through this construction we make sure
that we use only the information which is actually provided by the ND
measurements.

\section{A far detector-only setup with effective systematics}
\label{app:effectiveFD}

\newcommand{\FDeff}{$\mathrm{FD^{eff}}$}

Previous sensitivity studies (such as for example the ones in
Refs.~\cite{Group:2007kx, Huber:2002mx,Campagne:2006yx}) do not
include a ND in the simulation, and some values for systematical
uncertainties are adopted, which are assumed to implicitly encode
information from the ND. In this appendix we examine in which way such
choices for systematical errors should be interpreted. We calculate
the sensitivity to CPV for T2HK for a FD-only configuration (denoted
by \FDeff), by using exactly the same detector simulation and
backgrounds as before, but we include only four independent effective
systematical uncertainties: the normalisations of the appearance
signal and the normalisations of the total background, both for
neutrinos and anti-neutrinos ($\sigma^\mathrm{sig}_\nu$,
$\sigma^\mathrm{sig}_{\anu}$, $\sigma^\mathrm{bg}_\nu$,
$\sigma^\mathrm{bg}_{\anu}$). According to the discussion in
Sec.~\ref{sec:qualitative} one expects that $\sigma^\mathrm{bg}_\nu$
and $\sigma^\mathrm{bg}_{\anu}$ will be relevant for the sensitivity
at small $\theta_{13}$, whereas $\sigma^\mathrm{sig}_\nu$ and
$\sigma^\mathrm{sig}_{\anu}$ will dominate the sensitivity at large
$\theta_{13}$, and this is indeed the behaviour we find.

\begin{figure}[t] \centering 
    \includegraphics[width=0.6\textwidth]{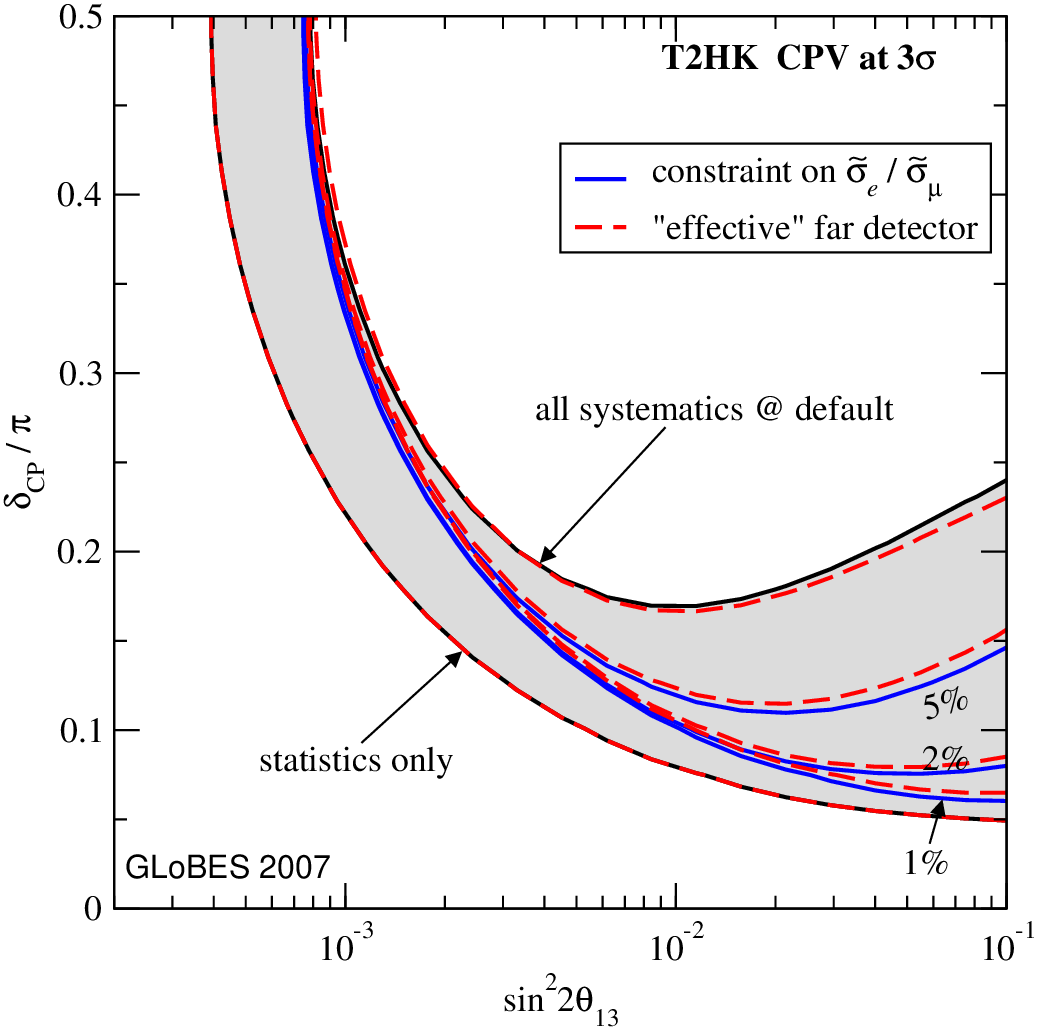} 
    \mycaption{\label{fig:FDeff} Comparison of the effective FD
    description (dashed curves) with the full ND/FD setup. The shaded
    region, as well as the blue curves correspond to our standard
    simulation and are identical to the corresponding curves in
    Fig.~\ref{fig:T2HK-CPV}.}
\end{figure}

In Fig.~\ref{fig:FDeff} we compare this \FDeff\ simulation with the full
ND/FD configuration. As it must be, the pure statistics sensitivities
are identical. It turns out that the ND/FD sensitivity with all
systematics at the default values according to Fig.~\ref{fig:pulls} is
reproduced rather accurately by \FDeff\ for the following choice of
systematics:
\begin{equation}
\sigma^\mathrm{sig}_\nu = \sigma^\mathrm{sig}_{\anu} = 10\% 
\,,\quad
\sigma^\mathrm{bg}_\nu = \sigma^\mathrm{bg}_{\anu} = 3.5\% \,.
\end{equation}
We conclude from these numbers that our specific implementation of the
ND provides a 3.5\% measurement of the background, whereas the
effective error on the signal turns out to be 10\%. To reproduce the
curves corresponding to a constraint on the
$\tilde\sigma_{\nu_e}/\tilde\sigma_{\nu_\nu}$ ratio at $x\%$ in the
full ND/FD case, one simply has to set $\sigma^\mathrm{sig}_\nu =
\sigma^\mathrm{sig}_{\anu} = x\%$ for \FDeff. This confirms the
arguments given in Sec.~\ref{sec:qualitative}, that the error on the
ratio $\tilde\sigma_{\nu_e}/\tilde\sigma_{\nu_\nu}$ directly
translates into an error on the appearance signal.

The \FDeff\ calculations can be performed with a standard {\sf GLoBES}
analysis. A glb-file for this effective FD simulation for T2HK is
available at~\cite{globesweb}.


\bibliographystyle{apsrev}
\bibliography{./syst}

\end{document}